\documentclass[english,12pt,pre,unsortedaddress,nofootinbib]{revtex4-1}
\usepackage{ae,aecompl}
\usepackage[T1]{fontenc}
\usepackage[latin9]{inputenc}
\usepackage{float}
\usepackage{amsmath,amssymb}
\usepackage{array}
\usepackage{graphicx}
\usepackage{color}
\usepackage{subcaption}
\usepackage{ragged2e}
\DeclareCaptionJustification{justified}{\justifying}
\usepackage{enumitem}
\usepackage{setspace}
\makeatletter

\floatstyle{ruled}
\newfloat{algorithm}{H}{loa}
\providecommand{\algorithmname}{Algorithm}
\floatname{algorithm}{\protect\algorithmname}

\makeatother

\usepackage{babel}
\begin{document}
\global\long\def\V#1{\boldsymbol{#1}}
\global\long\def\M#1{\boldsymbol{#1}}
\global\long\def\Set#1{\mathbb{#1}}

\global\long\def\D#1{\Delta#1}
\global\long\def\d#1{\delta#1}

\global\long\def\norm#1{\left\Vert #1\right\Vert }
\global\long\def\abs#1{\left|#1\right|}

\global\long\def\grad{\M{\nabla}}
\global\long\def\avv#1{\langle#1\rangle}
\global\long\def\av#1{\left\langle #1\right\rangle }

\global\long\def\P{\mathcal{P}}

\global\long\def\ki{k}
\global\long\def\wi{\omega}

\global\long\def\slip{\breve{\V u}}

\global\long\def\bu{\V u}
 \global\long\def\bv{\V v}
 \global\long\def\br{\V r}

\global\long\def\sM#1{\M{\mathcal{#1}}}
\global\long\def\fM#1{\M{\mathfrak{#1}}}
\global\long\def\Mob{\sM M}
\global\long\def\J{\sM J}
\global\long\def\S{\sM S}
\global\long\def\L{\sM L}
\global\long\def\R{\sM R}

\global\long\def\N{\sM N}
\global\long\def\K{\sM K}
\global\long\def\slipN{\breve{\N}}


\global\long\def\Lub{\D{\sM{R}}}
\global\long\def\LubPair{\Lub^{\text{pair}}}
\global\long\def\LubWall{\Lub^{\text{wall}}}

\global\long\def\Rsup{\sM{R}^{\text{sup}}}
\global\long\def\Rsuplub{\Rsup_{\text{lub}}}
\global\long\def\RsupRPY{\Rsup_{\text{RPY}}}

\global\long\def\Rpair{\sM{R}^{\text{pair}}}
\global\long\def\RpairRPY{\Rpair_\text{{RPY}}}
\global\long\def\Rpairlub{\Rpair_\text{{lub}}}

\global\long\def\Rwall{\sM{R}^{\text{wall}}}
\global\long\def\RwallRPY{\Rwall_{\text{RPY}}}
\global\long\def\Rwalllub{\Rwall_{\text{lub}}}


\global\long\def\MBlock{\Mob_{\text{free}}}
\global\long\def\aN{\widetilde{\N}}
\global\long\def\aK{\widetilde{\K}}
\global\long\def\aMob{\widetilde{\Mob}}
\global\long\def\fMob{\fM M}
\global\long\def\fR{\fM R}

\global\long\def\epsN{\overline{\N}}

\global\long\def\epsM{\overline{\Mob}}

\global\long\def\slipW{\breve{\V W}}
\global\long\def\rot{\M{\Psi}}
\global\long\def\Rot{\M{\Xi}}

\global\long\def\rhat{\hat{\V{r}}}
\global\long\def\zhat{\hat{\V{z}}}

\global\long\def\eqd{\overset{d}{=}}

\newcommand{\modified}[1]{#1}
\newcommand{\deleted}[1]{}

\title{\modified{Driven} dynamics in dense suspensions of microrollers}

\author{Brennan Sprinkle}
\affiliation{Courant Institute of Mathematical Sciences, New York University,
New York, NY 10012}

\author{Ernest B. van der Wee}
\affiliation{Department of Physics and Astronomy, Northwestern University, Evanston,
IL 60208}

\author{Yixiang Luo}
\affiliation{Department of Mathematics, UC Berkeley, Berkeley CA 94720}
\affiliation{Courant Institute of Mathematical Sciences, New York University,
New York, NY 10012}

\author{Michelle Driscoll}
\email{michelle.driscoll@northwestern.edu}
\affiliation{Department of Physics and Astronomy, Northwestern University, Evanston,
IL 60208}

\author{Aleksandar Donev}
\email{donev@courant.nyu.edu}
\affiliation{Courant Institute of Mathematical Sciences, New York University, New York, NY 10012}

\begin{abstract}
We perform detailed computational and experimental measurements of the \modified{driven} dynamics of a dense, uniform suspension of sedimented microrollers driven by a magnetic field rotating around an axis parallel to the floor. We develop a lubrication-corrected Brownian Dynamics method for dense suspensions of driven colloids sedimented above a bottom wall. The numerical method adds lubrication friction between nearby pairs of particles, as well as particles and the bottom wall, to a minimally-resolved model of the far-field hydrodynamic interactions. Our experiments combine fluorescent labeling with particle tracking to trace the trajectories of individual particles in a dense suspension, and to measure their propulsion velocities. Previous computational studies [B. Sprinkle \emph{et al.}, J. Chem. Phys., 147, 244103, 2017] predicted that at sufficiently high densities a uniform suspension of microrollers separates into two layers, a slow monolayer right above the wall, and a fast layer on top of the bottom layer. Here we verify this prediction, showing good quantitative agreement between the bimodal distribution of particle velocities predicted by the lubrication-corrected Brownian Dynamics and those measured in the experiments. The computational method accurately predicts the rate at which particles are observed to switch between the slow and fast layers in the experiments. We also use our numerical method to demonstrate the important role that pairwise lubrication plays in motility-induced phase separation in dense monolayers of colloidal microrollers, as recently suggested for suspensions of Quincke rollers [D. Geyer \emph{et al.}, Physical Review X, 9(3), 031043, 2019].
\end{abstract}

\maketitle

\section{Introduction}

Driven suspensions of colloidal microrollers \cite{Rollers_NaturePhys,MagneticRollers,BrownianMultiblobSuspensions} provide a simple but rich test-bed to explore emergent, collective hydrodynamic phenomena in active systems. The magnetic microrollers studied in this work are spherical colloids with an embedded ferromagnetic cube of hematite, which gives the particles a permanent magnetic moment that is sufficiently strong to drive them with an external magnetic field, but weak enough \emph{not} to induce significant inter-particle magnetic interactions \cite{Rollers_NaturePhys}. A rotating magnetic field can  be used to spin the particles in phase with the applied field. When the colloids are sedimented above a bottom wall and the magnetic field rotates around an axis parallel to the floor, the broken symmetry converts their angular velocity into linear velocity \cite{NearWallSphereMobility}, creating an active suspension \cite{Rollers_NaturePhys}. The collective flows generated in dense suspensions increase the active velocity and lead to unusual dynamics, such as the formation of stable self-propelled clusters of microrollers termed ``critters'' in \cite{Rollers_NaturePhys}. 

Some of us showed in \cite{MagneticRollers} that thermal fluctuations are crucial to the dynamics of microrollers as they set a characteristic height of the particles above the wall, which in turn controls the size of the critters. In subsequent work \cite{BrownianMultiblobSuspensions}, some of us used numerical simulations to predict that sufficiently dense, uniform suspensions of microrollers will self separate into two groups: one group of particles which moves slowly and stays close to the wall, and another which lies above the first and travels much faster. In this work, we provide the first experimental validation of this type of active particle separation, and introduce a lubrication-corrected Brownian Dynamics numerical method to model the experiments. Our method is simple and efficient by virtue of minimally resolving the far-field hydrodynamics, yet, as we show, provides sufficient quantitative accuracy to reproduce our experimental results.

Previous studies of the driven microroller suspensions obtained good qualitative agreement between simulations and experiments \cite{Rollers_NaturePhys,NonlocalShocks_Rollers,MagneticRollers}, however, quantitative agreement was lacking for two reasons. First, the minimally-resolved hydrodynamics based on the Rotne-Prager-Yamakawa (RPY) approximation did not correctly account for near-field hydrodynamics. Second, the experiments used Particle Image Velocimetry (PIV) to measure the mean suspension velocity, and PIV may give wrong results when there are height-separated slow and fast particles. Specifically, in \cite{NonlocalShocks_Rollers} the dispersion relationship of a uniform suspension of microrollers was measured experimentally and predicted by a continuum model based on the RPY tensor, and it was found that "The mean suspension velocity obtained from the continuum model ... overestimates the one measured in the experiments by a factor of around 4-5."

The lubrication-corrected Brownian Dynamics (BD) method we present here adds lubrication corrections to the minimally-resolved BD method described in \cite{MagneticRollers} in order to enable more accurate modeling of densely-packed Brownian suspensions of spherical colloids. This allows us to interrogate dense, nearly two-dimensional suspensions, and to make quantitative predictions that can directly be compared to experiments. We also report here new experimental results on the driven dynamics of uniform suspensions of microrollers. We fluorescently label only a small subset of the particles in order to enable particle tracking in  the plane parallel to the wall, even in dense suspensions and in the presence of multiple layers of particles. This allows us to experimentally measure the distribution of active velocities, as well as to measure dynamical correlation functions for a single particle.

Lubrication corrections were originally introduced in Stokesian Dynamics (SD) \cite{SD_Suspensions}, but have since been incorporated in a variety of related methods for Stokesian suspensions. The key idea is to account for the near-field pairwise lubrication forces in the resistance formulation, and for the far-field hydrodynamic interactions in the mobility formulation, and combine the two to give a lubrication-corrected mobility matrix. The far-field approximation itself can be obtained by a variety of numerical techniques, ranging from the minimially-resolved RPY mobility we use here, through multipole expansions with higher-order multipoles \cite{HydroMultipole_Ladd_Lubrication,HYDROMULTIPOLE_Wall,ForceCoupling_Lubrication,SE_Multiblob_SD}, to boundary integral methods \cite{BrownianMultiblobSuspensions,BoundaryIntegralGalerkin}. The pairwise lubrication approximation is not always accurate \cite{LubricationLamb_3spheres} and the accuracy \emph{cannot} be controlled \emph{a priori}. Nevertheless, lubrication corrections provide a means of substantially increasing the hydrodynamic accuracy for dense suspensions, while keeping the computational cost small enough to enable practical large-scale and long-time simulations.

Recently, Fiore and Swan developed a fast Stokesian Dynamics method that can include Brownian motion with a cost essentially linear in the number of particles \cite{SE_Multiblob_SD}. To this end they use a combination of sophisticated numerical linear algebra and the positively split Ewald method of \cite{SpectralSD,SpectralRPY} to simultaneously account for the Brownian forces as well as the lubrication corrections. The method we present in this work to simulate Brownian particle suspensions is similar to the method developed by Fiore and Swan in \cite{SE_Multiblob_SD}, with a few important differences. Firstly, the work in \cite{SE_Multiblob_SD} was tailored to periodic (bulk) suspensions of particles in 3D, while ours is tailored to suspensions above a bottom wall. The inclusion of a bottom wall requires applying lubrication corrections when particles approach the wall, and the hydrodynamic screening with the bottom wall makes the far-field mobility matrix better conditioned, simplifying the linear algebra required. Secondly, since we do not study rheology, we omit the stresslet constraints, which greatly improves the efficiency without sacrificing the improvement in accuracy due to the lubrication corrections \footnote{Mathematically, the torque and stresslet moments enter at the same level of the multipole hierarchy and should thus, in principle, be both included or both omitted. However, we show here empirically that the stresslets can be omitted in practice for the types of problems we study here.}. Our minimally-resolved approach allows for the design of a novel preconditioning strategy, as well as a novel temporal integration scheme which achieves greater temporal accuracy than the scheme used by Fiore and Swan, while also reducing the computational cost.

In this paper we develop a minimally-resolved BD method for suspensions above a bottom wall that incorporates lubrication corrections, and apply the method to simulating suspensions of microrollers. In section \ref{sec:detLub}, we describe in detail a deterministic method to account for near-field lubrication corrections, and outline the necessary modifications required to account for the confinement by a bottom wall. In section \ref{sec:BD} we account for thermal fluctuations and describe an efficient and accurate lubrication-corrected BD method for driven suspensions above a bottom wall, including a novel predictor-corrector temporal integration scheme.  

Section \ref{sec:MagRoll} revisits the active dynamics of a uniform suspension of magnetic rollers above a bottom wall. Some of us previously used the rigid multiblob method to predict a bimodal distribution in the particles' velocities, caused by the bimodal distribution of their heights above the wall \cite{BrownianMultiblobSuspensions}. We reproduce these predictions here using the simpler and more efficient lubrication-corrected BD method, and confirm the bimodality experimentally by using particle tracking. By comparing results between experiments and simulation, we demonstrate that modeling the propulsive mechanism of the microrollers using a constrained angular velocity is more physically accurate than using a constant applied torque, as was done in prior work \cite{MagneticRollers,BrownianMultiblobSuspensions}. To this end, we design a novel preconditioned iterative method to efficiently constrain the angular velocity of the microrollers to a prescribed value.

In \cite{MIPS_Quincke_Bartolo}, Geyer \emph{et al.} argue that active Quincke rollers densely packed above a bottom wall will, at sufficiently large densities, slow down and even crystalize in an almost immobile solid phase, because of the pairwise lubrication friction between nearly touching colloids. Inspired by this work, in section \ref{sec:Sheet} we use our lubrication-corrected BD method to study the collective dynamics of a sheet of microrollers constrained to a fixed height just above the bottom wall. We study the dependence of the mean (collective) velocity on the in-plane packing fraction, and show that this trend is qualitatively different when prescribing activity using a constant applied torque versus prescribing a constant angular velocity.

\section{\label{sec:detLub}Lubrication Corrections}

In this work, we are concerned with simulating the dynamics of $N$ spherical particles with uniform radii $a$ of at most a few microns. This length scale is small enough to consider the effect of fluid inertia negligible and to treat the hydrodynamics of the particle suspension using the Stokes equations with no-slip conditions on the surfaces of the particles as well as the surface of the bottom wall. Furthermore, the Brownian motion due to thermal fluctuations of the fluid should not be neglected. Nevertheless, we will briefly ignore fluctuations in this section, and return to Brownian motion in Section \ref{sec:BD}.

The linearity of the Stokes equations ensures that we can write the translational velocities $\V u_i$ and angular velocities $\V \omega_i$ of all particles $1 \leq i \leq N$ in terms of the forces $\V f_i$ and torques $\V \tau_i$ applied to the particles, using the hydrodynamic \emph{mobility matrix} $\fMob$,
\begin{equation} \label{eq:MF}
\V U = \fMob \V F,
\end{equation}
where the vector of linear and angular velocities is $\V U = \left[\V u_1, \V \omega_1, \V u_2, \V \omega_2, \cdots, \V u_N, \V \omega_N\right]^{T}$, and the vector of applied forces and torques is $\V F = \left[\V f_1,\V \tau_1, \V f_2, \V \tau_2, \cdots, \V f_N, \V \tau_N\right]^{T}$ (where the superscript $T$ denotes a transpose). The inverse of the mobility matrix is the \emph{resistance matrix} $\fR = \fMob^{-1}$. The mobility and resistance matrices will in general depend of the positions and orientations of all of the particles $\V{Q}=\left[\V{q}_1,\cdots,\V{q}_N\right]^{T}$, though we will often omit the explicit dependence for simplicity of notation. Because the particles we consider are spherical, the mobility does not depend on their orientation, however, we explicitly track and evolve the orientation of every particle in our numerical methods. 

Computing the action of the true mobility matrix (i.e., solving the mobility problem) with high accuracy is very expensive for many-particle suspensions even at moderate densities \cite{Stokes3D_FMM,SphericalHarmonics_Yan}. A commonly used approximation to the hydrodynamic mobility is a pairwise approximation $\fMob \approx \Mob_{\text{RPY}}$ based on the Rotne-Prager-Yamakawa (RPY) tensor \cite{RotnePrager,RPY_Yamakawa,RPY_Shear_Wall}. This regularized form of the mobility is sufficiently accurate in resolving hydrodynamic interactions if particles are well separated, and ensures that the mobility matrix is symmetric positive semidefinite \cite{RPY_Shear_Wall}; this is an essential property when including Brownian motion. Originally the RPY tensor was formulated for particle suspensions in free space, but Swan and Brady give a modified Rotne-Prager-Blake form which accounts for an unbounded (in the transverse directions) bottom wall in \cite{StokesianDynamics_Wall}. The wall corrections from \cite{StokesianDynamics_Wall} can be combined with the overlapping corrections as described in \cite{RPY_Shear_Wall} to give analytical expressions for the elements of $\Mob \equiv \Mob_{\text{RPY}}$, as described in more detail in \cite{MagneticRollers}. Efficiently computing $\Mob_{\text{RPY}} \V{F}$ in time approximately linear in the number of particles is not trivial but is possible, including for systems that are periodic in some of the transverse directions, using Fast Multipole Methods (FMMs) \cite{FMM_wall} or the Fast Fourier Transform (FFTs) \cite{SpectralEwald_Wall}. Here we rely on Graphical Processing Units (GPUs) to dramatically accelerate the direct (quadratic cost) computation, but more advanced methods can be substituted depending on the available software, hardware, and the number of particles.

It is important to note that the Stokesian Dynamics formulation \cite{SD_Suspensions,StokesianDynamics_Wall,BrownianDynamics_OrderNlogN,SE_Multiblob_SD,SD_LubOnly} also accounts for shear and stresslets but we will omit the stresslet blocks in the spirit of a minimally-resolved approach; the reader can consult the recent work of Fiore and Swan \cite{SE_Multiblob_SD} for how to efficiently include stresslet terms in $\Mob$, at the expense of increased computational complexity. This makes our method much simpler to implement in the presence of a wall and also more efficient, but note that rheological properties cannot be studied without accounting for the particle stresslets \footnote{Note that omitting the far-field mobility would make the method even more efficient but would not be able to reproduce the collectively-generated active flows studied here, and can lead to unphysical results in general \cite{SD_LubOnly}.}. We study the deterministic accuracy of our approach in Appendix \ref{app:Tetrahedron}, and find that even without stresslets the lubrication corrections lead to a rather accurate mobility matrix over a range of distances.

The RPY mobility inaccurately resolves near-field hydrodynamic interactions and cannot be used for dense suspensions if quantitative accuracy is desired.  The essential motivation behind the lubrication corrections used in Stokesian Dynamics \cite{SD_Suspensions} is to maintain the desirable properties of the RPY tensor in the far field but correct for its poor near-field hydrodynamic resolution. The approach is to add a local pairwise correction to the RPY resistance matrix $\R =\Mob^{-1}\equiv \R_{\text{RPY}}$ for all pairs of surfaces (i.e two spheres or a sphere and the wall) which are sufficiently close. The lubrication correction resistance matrix $\Rsuplub$ is assembled from accurate resistance matrices for each pair of nearly touching surfaces (i.e., two spheres, or a sphere and a wall). The corrections are applied to the resistance matrix rather than the mobility matrix because near-field hydrodynamic interactions are approximately pairwise additive in resistance form, unlike far-field interactions which are approximately pairwise additive in mobility form. In analogy with classical asymptotic methods, the full lubrication-corrected mobility $\epsM$ is constructed by subtracting off the ``common part'' $\RsupRPY$, i.e., the overlapping near-field contributions between $\R$ and $\Rsuplub$, giving the \emph{lubrication-corrected mobility}
\begin{equation} \label{eq:epsM1}
\fMob \approx \epsM = \left[\R + \Rsuplub - \RsupRPY  \right]^{-1} 
\end{equation}
Here $\RsupRPY$ is assembled from pairwise RPY resistance tensors for the same pairs of nearby surfaces included when constructing $\Rsuplub$.

In this section we detail how to simulate driven particle suspensions above a wall, accounting for lubrication corrections, but neglecting thermal fluctuations. Specifically, we first describe how lubrication corrections are applied to the RPY hydrodynamic mobility $\Mob$ in the presence of a bottom wall. We then describe a preconditioned Krylov method to apply the lubrication-corrected mobility to a vector of applied forces and torques. While here we focus on deterministic dynamics, special care will be taken to ensure that Brownian motion can be included, i.e., that the lubrication-corrected mobility is positive definite. While our method is closely-related to the fast Stokesian Dynamics method recently presented by Fiore and Swan \cite{SE_Multiblob_SD} for periodic suspensions, there are several differences that we detail in this section. Specifically, we consider here suspensions sedimented above a bottom wall, exclude the stresslet corrections since we are not concerned with rheology, and develop a different preconditioner.

\subsection{\label{sec:DR}Lubrication Corrected Mobility}

The lubrication-corrected mobility $\epsM$ defined in equation \eqref{eq:epsM1} can be restated as \cite{libStokes,SE_Multiblob_SD}
\begin{eqnarray}
\epsM & = & \left[\Mob^{-1}+\Lub\right]^{-1} \label{eq:LubN}  \\
 & = & \Mob \cdot\left[\M I+\Lub \cdot \Mob \right]^{-1}\nonumber
\end{eqnarray}
where $\Lub=\Rsuplub-\RsupRPY$ is the lubrication
correction for the resistance matrix. The basic idea \cite{SD_Suspensions} is to subtract off 
the RPY mobility for all nearby pairs of surfaces, and replace
it with an exact analytic formula, while maintaining the long-ranged hydrodynamics using the RPY mobility/resistance.

Both $\Rsuplub$ and $\RsupRPY$ take the general form of a pairwise-additive resistance matrix $\Rsup$, which is assembled by summing appropriate blocks of the symmetric, pair-resistance matrices between particles $i$ and $j$, 
\begin{equation}
\Rpair \left(\V{q}_i,\V{q}_j\right)  = 
\begin{bmatrix}
\Rpair_{\text{self}}\left(\V{q}_i,\V{q}_j\right) & \Rpair_{\text{couple}}\left(\V{q}_i,\V{q}_j\right) \\
\Rpair_{\text{couple}}\left(\V{q}_j,\V{q}_i\right) & \Rpair_{\text{self}}\left(\V{q}_i,\V{q}_j\right)
\end{bmatrix}.
\end{equation}
Treating the wall as a surface which hydrodynamically interacts with each particle through a pair-resistance matrix $\Rwall\left(\V{q}_i\right)$, $\Rsup$ is assembled as
\begin{equation} \label{grandLub}
\Rsup = 
\begin{bmatrix}
\displaystyle \sum_{j \neq 1} \Rpair_{\text{self}}\left(\V{q}_1,\V{q}_j\right) + \Rwall \left(\V{q}_1\right) & \Rpair_{\text{couple}}\left(\V{q}_1,\V{q}_2\right) & \cdots  \\
\Rpair_{\text{couple}}\left(\V{q}_2,\V{q}_1\right) & \displaystyle \sum_{j \neq 2} \Rpair_{\text{self}}\left(\V{q}_2,\V{q}_j\right) + \Rwall \left(\V{q}_2\right) & \cdots  \\
\vdots & \ddots & \ddots
\end{bmatrix},
\end{equation}
where $\Rpair_{\text{couple}}\left(\V{q}_i,\V{q}_j\right) = \left(\Rpair_{\text{couple}}\left(\V{q}_j,\V{q}_i\right)\right)^{T}$.

\subsection{Computing $\Lub$}

Each block of $\Rpair$, either $\Rpairlub$ or $\RpairRPY$, can be expressed in terms of coefficients which depend on the dimensionless gap between the surfaces of the spheres,   
\begin{equation} \label{Rpair}
\Rpair_{s,c}  \left(\V{q}_i,\V{q}_j\right)=
\begin{bmatrix}
X^{tt}_{s,c}\left(\epsilon_r\right)\rhat \rhat^{T}+Y^{tt}_{s,c}\left(\epsilon_r\right)\left(\M{I}-\rhat \rhat^{T} \right) & -Y^{tr}_{s,c}\left(\epsilon_r\right) \rhat\times \\
Y^{tr}_{s,c}\left(\epsilon_r\right) \rhat\times & X^{rr}_{s,c}\left(\epsilon_r\right)\rhat \rhat^{T}+Y^{rr}_{s,c}\left(\epsilon_r\right)\left(\M{I}-\rhat \rhat^{T} \right)
\end{bmatrix},
\end{equation}
where $a$ is the radius of the particles,
\[
\epsilon_r = \frac{||\V{q}_j - \V{q}_i||}{a} - 2, \ \ \rhat = \frac{\V{q}_j - \V{q}_i}{||\V{q}_j - \V{q}_i||},
\]
and $s,c$ indicates whether this is the `self' or the `couple' block. In \eqref{Rpair}, the matrix $\rhat\times$ represents a cross product by $\rhat$ and the superscripts on the coefficients denote the type of the block, e.g $Y^{tr}_{s,c}$ denotes that this is the coefficient of the translation `t' and rotation `r' coupling block.
Because the coefficients of $\Rpair$ decay as $\epsilon_r$ grows, we set a cutoff distance, $\epsilon_r^{cut}$ such that $\Rpair = \M 0$ for $\epsilon_r > \epsilon_r^{cut}$. A smaller value for $\epsilon_r^{cut}$ ensures that $\Rsup$ is more sparse and therefore easier to construct and apply, but this, of course, comes at the cost of reduced accuracy. In this work we have found that $\epsilon_r^{cut} = 2.5$ strikes a good balance, and so we use this value throughout.

Wall corrections to the self resistance, either $\Rwalllub$ or $\RwallRPY$, have a similar form to $\Rpair_{s}$ but the coefficients depend instead on the dimensionless wall separation $\epsilon_h$, such that
\begin{equation} \label{RwallXY}
\Rwall \left(\V{q}_i\right) =
\begin{bmatrix}
X^{tt}_{\text{wall}}\left(\epsilon_h\right)\zhat \zhat^{T}+Y^{tt}_{\text{wall}}\left(\epsilon_h\right)\left(\M{I}-\zhat \zhat^{T} \right) & -Y^{tr}_{\text{wall}}\left(\epsilon_h\right) \zhat\times \\
Y^{tr}_{\text{wall}}\left(\epsilon_h\right) \zhat\times & X^{rr}_{\text{wall}}\left(\epsilon_h\right)\zhat \zhat^{T}+Y^{rr}_{\text{wall}}\left(\epsilon_h\right)\left(\M{I}-\zhat \zhat^{T} \right),
\end{bmatrix}
\end{equation}
where $\zhat$ is the unit vector perpendicular to the wall, and
\[
\epsilon_h = \frac{\V{q}_i \cdot \zhat}{a} - 1.
\]
Unlike the pair corrections between nearby particles which have a cutoff distance, we will always apply wall corrections to each particle. This ensures that the diagonal blocks of $\Rsup$ are never exactly zero for particles reasonably close to the bottom wall --- a feature which we will find useful for designing efficient linear solvers in section \ref{sec:linsys}.

Given accurate values or formulas for the coefficients of $\Rpairlub$ and $\Rwalllub$ when $\epsilon_r$ and/or $\epsilon_h$ are small, we may form a pairwise, wall--corrected, nearfield resistance matrix $\Rsuplub$ using \eqref{grandLub}.  Analytical or semi-analytical formulas for $\Rpairlub$ and $\Rwalllub$ are summarized in appendix \ref{apdx:RPRW}.
\modified{For very small values of the dimensionless gap $\epsilon_r$, the resistance functions appearing in \eqref{Rpair} are known to have universal asymptotic forms including also for spheres of different radii. Since the case of a flat wall and a sphere is the limit of one of the two spheres (the `wall') being infinitely larger than the other \cite{SphericalHarmonicImages_Wall}, the same applies for the resistance functions of $\epsilon_h$. Asymptotic expansions of resistance functions typically involve constants, $1/\epsilon$, $\ln \epsilon$, and $\epsilon \ln \epsilon$ terms, see \cite{SpheresStokes_JeffreyOnishi,HYDROMULTIPOLE_Theory,MicrohydrodynamicsBook} for more details, and these asymptotic forms of the resistance are sometimes referred to as ``lubrication friction''. Here we instead use the term ``lubrication'' to refer more generally to near-field hydrodynamics not included in the far-field approximation.}
As detailed in appendix \ref{apdx:RPRW}, when no known analytical formula is sufficiently accurate, we use the rigid multiblob method \cite{RigidMultiblobs,BrownianMultiblobSuspensions} to compute a numerical approximation.

\subsection{A positive definite form for $\Lub$}

In order to include Brownian motion, it is important that $\Lub$ be positive semi-definite, ensuring that a `square root' $\left(\Lub\right)^{1/2}$ exists.
The resistance correction $\Lub$ will be positive semi-definite if each pairwise block is. This is empirically known to be true in the absence of a bottom wall when stresslets are included, as discussed in more detail \cite{SE_Multiblob_SD}. We are, however, not aware of a mathematical proof, or any prior studies investigating this for a sphere and a bottom wall.

Numerically, we find that in the presence of a bottom wall, $\Lub$ can
have small negative eigenvalues. These small eigenvalues come directly
from the wall contribution to $\Lub$ which we term $\LubWall$. For each particle whose height $h \gtrsim 1.5a$,
$\LubWall$ has at least one small negative
eigenvalue 
caused by discretization error in the rigid multiblob method \cite{RigidMultiblobs} we use to calculate $\Rwalllub$, for lack of an exact method. 
A simple remedy is to diagonalize $\LubWall$ and replace the spurious negative eigenvalues by 0 to form $\LubWall_{\lambda>0}$. We also need to remove the negative eigenvalues in $\Rwalllub$, which we need for the preconditioner described in section \ref{sec:linsys},
\[
 \left(\Rwalllub \right)_{\lambda>0} = \LubWall_{\lambda>0} + \RwallRPY.
\]
This construction ensures that $\Lub=\LubPair + \LubWall_{\lambda>0}$ is positive semi-definite.

\subsection{Linear Algebra} \label{sec:linsys}

Given a vector of applied forces and torques on a suspension of particles $\V{F}$, we need an efficient method to apply the lubrication-corrected mobility $\epsM$ to find the resulting linear and angular velocities $\V{U}=\epsM\V{F}$,    
\begin{align}
\V U & =\left[\M I+\Mob\Lub\right]^{-1}\Mob\V F,\label{eq:OBDet} \\
     & =\Mob \left[\M I+\Lub\Mob \right]^{-1}\V F.\label{eq:OBDetalt}
\end{align}
We compute the action of either $\left[\M I+\Mob\Lub\right]^{-1}$ or $\left[\M I+\Lub\Mob \right]^{-1}$ on a vector using an efficient preconditioned Krylov method.

If we wish to use equation \eqref{eq:OBDet} to apply $\epsM$ efficiently \footnote{A preconditioner for equation \eqref{eq:OBDetalt} can be developed through a similar method.}, we must solve a system of the form
\begin{equation} \label{eq:LubSolve1}
 \left[\M I+\Mob\Lub\right] \V{x} = \V{b}.
\end{equation}
To develop a preconditioner for equation \eqref{eq:LubSolve1}, we ignore the far-field hydrodynamics and approximate $\Mob \approx \left(\RsupRPY\right)^{-1}$, giving
\begin{align} 
  \V{x} &= \left[\M I+\Mob\Lub\right]^{-1} \V{b} \approx \left[\M I+\left(\RsupRPY\right)^{-1} \Lub \right]^{-1} \V{b} \\
	&= \left[\M I+\left(\RsupRPY\right)^{-1} \left(\Rsuplub - \RsupRPY \right) \right]^{-1} \V{b}   \\
	&= \left(\left(\Rsuplub\right)^{-1} \RsupRPY\right) \V{b} = \M{P}_1 \V{b}. \label{eq:PC1}
\end{align}
We compute $\left(\Rsuplub\right)^{-1}$ using the super-nodal Cholesky solver provided in the CHOLMOD package \cite{cholmod}, which is very efficient due to the quasi two--dimensional nature of sedimented suspensions. Note that an incomplete Cholesky decomposition could also be used here as was done in \cite{SE_Multiblob_SD}. In all of the numerical experiments performed here, both Cholesky solves and Cholesky factorizations using CHOLMOD take substantially less time than a single multiplication by the RPY mobility tensor $\Mob$.

A different preconditioner was obtained in \cite{SE_Multiblob_SD} by approximating $\Mob$ by a block diagonal matrix, $\MBlock$, where each block is given by the freespace mobility of a single sphere
\[
 \left[\MBlock\right]_{ii} = \begin{bmatrix} \frac{1}{6 \pi \eta a} \M{I} & \M{0} \\
 \M{0} & \frac{1}{8 \pi \eta a^3} \M{I}
 \end{bmatrix}.
\]
The resulting preconditioner can be stated as 
\begin{equation} \label{eq:PCbd}
\V{x} \approx \M{P}_{2} \V{b} = \left(\M I+\MBlock \Lub\right)^{-1} \V{b},  
\end{equation}
where $\left(\M I+\MBlock \Lub\right)^{-1}$ can be efficiently applied using a super-nodal Cholesky solver, as for $\M{P}_{1}$. We show in appendix \ref{sec:GMRES} that for many cases the preconditioner $\M{P}_{2}$ performs comparably to $\M{P}_{1}$, however there are some case where $\M{P}_{1}$ outperforms $\M{P}_{2}$, and thus we use $\M{P}_{1}$ in this work.

In some systems, a few particles can become isolated from the bulk and cause some numerical difficulty in the proposed preconditioner \eqref{eq:PC1}. We define isolated particles as those which are not close enough to the wall to provide a substantial wall correction to the diagonal block of  $\Rsuplub$ (we use $h > 4.5 a$ in this work as a cutoff height for possible isolated particles) nor are they close enough to other particles to contribute a pair correction to $\Rsuplub$. These particles not only lead to poor conditioning of $\Rsuplub$, but the presence of isolated particles makes $\RsupRPY$ a poor approximation to $\Mob^{-1}$. To remedy this, we introduce a modified identify matrix $\M{I}_{\text{iso}}$ which is zero everywhere but contains $6 \times 6$ identity blocks on the blocks of the diagonal corresponding to isolated particles. Isolated particles can be considered nearly in free space, hence we modify the preconditioner \eqref{eq:PC1} to simply not apply to these particles:
\begin{equation} \label{eq:PCmod}
\V{x} \approx \M{P}_{1} \V{b} = \left(\M{I} - \M{I}_{\text{iso}} \right) \left(\Rsuplub + \epsilon \MBlock^{-1} \right)^{-1} \left(\M{I} - \M{I}_{\text{iso}} \right) \RsupRPY \V{b} + \M{I}_{\text{iso}} \V{b}.
\end{equation}
Here we regularize $\Rsuplub$ by an amount proportional to the GMRES solver tolerance $\epsilon$.  

\subsection{Specified Rotational Motion} \label{sec:constrain}

In order to simulate experiments involving microrollers we need to impose a prescribed angular velocity rather than a prescribed torque. That is, we need to solve for the required linear velocities $\V{u}$ and torques $\V \tau$, given some applied forces $\V{f}$ on the particles and the desired angular velocity $\V \omega$. This can be stated mathematically by rearranging the mobility problem as
\[
\epsM \begin{bmatrix} \V{f} \\
\V \tau
\end{bmatrix} = \left(\M I + \Mob \Lub \right)^{-1} \Mob \begin{bmatrix} \V{f} \\
\V \tau
\end{bmatrix} = \begin{bmatrix} \V{u}\\
\V \omega
\end{bmatrix},
\]
as a linear system in the unknown quantities
\begin{equation} \label{eq:constO}
\Mob \begin{bmatrix} \V 0 \\
\V \tau
\end{bmatrix} - \left(\M I + \Mob \Lub \right) \begin{bmatrix} \V{u} \\
\V 0
\end{bmatrix} = \left(\M I + \Mob \Lub \right) \begin{bmatrix} \V 0 \\
\V \omega
\end{bmatrix} - \Mob \begin{bmatrix} \V{f} \\
\V 0
\end{bmatrix}
=
\begin{bmatrix} \V a \\
\V b
\end{bmatrix}.
\end{equation}
We solve \eqref{eq:constO} for $[\V{u}, \V \tau]^{T}$ using a preconditioned GMRES method. As a preconditioner, we will solve \eqref{eq:constO} using the block diagonal freespace approximation $\Mob \approx \MBlock$. This results in a sparse, decoupled system of equations of the form
\begin{align} 
-\left( \M I + \frac{1}{6 \pi \eta a} \Lub^{tt} \right) \V{u} &= \V{a} \label{eq:PCO1} \\
\frac{1}{8 \pi \eta a^3} \V \tau -\left( \M I + \frac{1}{8 \pi \eta a^3} \Lub^{rt} \right) \V{u} &= \V{b} \label{eq:PCO2}
\end{align}
where $\Lub^{tt},\Lub^{rt}$ are the translation--translation and rotation--translation coupling blocks of $\Lub$ respectively. Equation \eqref{eq:PCO1} can be solved efficiently for $\V{u}$ using CHOLMOD, and given $\V{u}$, equation \eqref{eq:PCO2} is trivial to solve for $\V \tau$.

\section{\label{sec:BD}Brownian Dynamics}

In this section we describe how to account for thermal fluctuations, i.e., Brownian motion. Given the positive-definite, lubrication-corrected mobility matrix $\epsM(\V{Q})$, the Ito overdamped Langevin equation 
\begin{equation} \label{eq:Lang}
\frac{d \V{Q}}{dt} = \V U = \epsM \V{F} + (k_B T) \ \partial_{\V{Q}} \cdot \epsM + \sqrt{2 k_B T} \ \epsM^{1/2} \sM{W}(t),
\end{equation}
governs the particle dynamics in the presence of thermal fluctuations. In the above, $T$ denotes the solvent temperature, $k_B$ is Boltzmann's constant, and $\sM{W}(t)$ is a collection of independent white noise processes. The last term involving $\epsM^{1/2}$ is the Brownian increment, and the second term involving $\partial_{\V{Q}} \cdot \epsM$ is the stochastic drift. Note that the first equality in \eqref{eq:Lang} is just a shorthand notation because representing orientations requires using quaternions; the precise statement of the stochastic dynamics for full particle configurations, including their orientations, requires a more cumbersome notation and treatment which is described in \cite{BrownianMultiBlobs,BrownianMultiblobSuspensions}.

There are several challenges in solving equation \eqref{eq:Lang} efficiently. We need an efficient way to compute the deterministic dynamics $\V U = \epsM \V{F}$ with lubrication corrections; we discussed this already in section \ref{sec:detLub}. In the presence of thermal fluctuations surface overlaps (particle-particle or particle-wall) may occur, in which case the mobility needs to be carefully modified and the overlap must be separated in such a way as to maintain detailed balance. The Brownian increment also needs to be sampled efficiently --- in section \ref{sec:BI} we describe an efficient method of splitting $\epsM^{1/2} \sM{W}$ into near and far fields which is similar to what has been done in \cite{SE_Multiblob_SD,BrownianDynamics_OrderNlogN}. The drift term is more challenging to efficiently calculate --- in section \ref{sec:SDE} we develop a novel time integration scheme for \eqref{eq:Lang} which captures this term accurately and with minimal computational effort; our scheme is more specialized and efficient than the more general scheme developed in \cite{SE_Multiblob_SD}.

\subsection{\label{sec:BI}Generating Brownian Velocities} 

In order to perform Brownian dynamics simulations we need a method to efficiently compute normalized \footnote{The scaled Brownian velocities have covariance $\left(2 k_B T/\D{t}\right) \, \epsM$, where $\D{t}$ is the time step size.} Brownian ``velocities'' $\V U_{s}$, which are a Gaussian random vector with mean zero and covariance $\epsM$. Following \cite{SE_Multiblob_SD}, we generate $\V{U}_s$ as 
\begin{equation}\label{eq:SBroot}
\V U_{s} = \epsM \left( \Lub^{1/2} \M{W}_{1} + \Mob^{-1/2} \M{W}_{2}  \right) = \left[\M I+\Mob\Lub\right]^{-1} \left( \Mob \Lub^{1/2} \M{W}_{1} + \Mob^{1/2} \M{W}_{2}  \right),
\end{equation}
where $\M{W}_{1}$ and $\M{W}_{2}$ are independent standard Gaussian random vectors. It is easy to confirm that $\V U_{s}$ has the correct covariance,
\begin{align}
 \av{\V U_{s}\V U_{s}^{T}} &= \epsM \Big( \Lub^{1/2}  \av{\M{W}_{1}\left(\M{W}_{1}\right)^{T}}\Lub^{T/2} + \Mob^{-1/2}  \av{\M{W}_{2}\left(\M{W}_{2}\right)^{T}} \Mob^{-T/2} \Big) \epsM \\ &= \epsM \left( \Lub + \Mob^{-1} \right) \epsM = \epsM.
\end{align}
To compact the notation, we will write
\begin{equation} \label{eq:Mh1}
\left( \Mob \Lub^{1/2} \M{W}_{1} + \Mob^{1/2} \M{W}_{2}  \right) \eqd  \left( \Mob \Lub \Mob  + \Mob \right)^{1/2} \M{W}_{1,2}, 
\end{equation}
where $\M{W}_{1,2}$ is a vector of i.i.d. standard Gaussian variables. Here the equality is in distribution since the first and second moments of the left and right hand sides match. For the same reason, we can write in more compact notation,
\begin{equation} \label{eq:Mh2}
\V U_{s} = \left[\M I+\Mob\Lub\right]^{-1} \left( \Mob \Lub \Mob  + \Mob \right)^{1/2} \M{W}_{1,2} = \epsM^{1/2} \M{W}_{1,2}, 
\end{equation}
which defines a ``square root'' of the lubrication-corrected mobility matrix suitable for efficient sampling of Brownian velocities/increments.

In equation \eqref{eq:SBroot}, the term $\Lub^{1/2} \M{W}_{1}$ can be efficiently generated by separately generating pairwise and diagonal blocks using independent random numbers \cite{StokesianDynamics_Brownian,BrownianDynamics_OrderNlogN}. We prefer to numerically compute $\Lub^{1/2}$ as a sparse Cholesky factor of $\Lub$ using CHOLMOD, as this is very efficient in the quasi-2D geometry considered here. The terms involving $\Mob^{1/2} \M{W}_{2}$ in \eqref{eq:SBroot} are computed using the Lanczos--like method of \cite{SquareRootKrylov}, as was done in \cite{MagneticRollers,BrownianMultiblobSuspensions}. The convergence of the Lanczos--like method in a modest number of iterations (independent of the number of particles) is demonstrated in \cite{MagneticRollers} for just the `trans--trans' coupling block of $\Mob$; we observe similar convergence properties when the rotation coupling blocks are included.

\subsection{Stochastic Time Integration} \label{sec:SDE}

In this section we describe a temporal integration scheme to simulate the stochastic dynamics \eqref{eq:Lang}. Algorithm \ref{alg:trap} summarizes our integration scheme, termed the `Stochastic Trapezoidal Split' scheme or STS scheme. The mechanism by which the STS scheme captures the thermal drift is similar to the Trapezoidal-Slip scheme introduced in \cite{BrownianMultiblobSuspensions} to simulate Brownian dynamics of rigid particles using the rigid multiblob method. Both trapezoidal schemes use a combination of random finite differences (RFD) \cite{MagneticRollers,BrownianMultiblobSuspensions} and a trapezoidal predictor-corrector scheme to capture the stochastic drift. One major advantage of the STS scheme is that it only requires two linear solves per time step, in contrast to the three required by the Trapezoidal-Slip scheme \cite{BrownianMultiblobSuspensions} and by the Euler-Maruyama scheme used in \cite{SE_Multiblob_SD}. The STS scheme therefore achieves the second order accuracy of an analogous deterministic scheme (by virtue of being a trapezoidal method) with only a small additional cost to include the Brownian dynamics. A public-domain implementation of the STS scheme for lubrication-corrected BD can be found on github at \url{https://github.com/stochasticHydroTools/RigidMultiblobsWall}. 

The STS scheme is so named because it takes advantage of a product rule splitting of the thermal drift term 
\begin{align}
\partial_{\V{Q}} \cdot \epsM &= \partial_{\V{Q}} \cdot \left( \left[\M I+\Mob\Lub\right]^{-1} \Mob \right) \\
&= \left[\M I+\Mob\Lub\right]^{-1} \left( \partial_{\V{Q}} \cdot \Mob \right) + \left( \partial_{\V{Q}} \left[\M I+\Mob\Lub\right]^{-1} \right) \colon \Mob. \label{eq:splitRFD}
\end{align}
The scheme uses the idea of random finite differences \cite{BrownianMultiBlobs,BrownianMultiblobSuspensions} to capture the first term of \eqref{eq:splitRFD} and the natural drift produced by the trapezoidal scheme to capture the second term. Specifically, we will compute the quantity $\partial_{\V{Q}} \cdot \Mob$ according to the RFD formula 
\begin{equation} \label{eq:RFD}
 \partial_{\V{Q}} \cdot \Mob \approx \frac{1}{\delta} \av{ \left[ \Mob\left(\V Q + \delta \V W^{D} \right)-\Mob\left(\V Q - \delta \V W^{D} \right) \right] \V W^{f \tau} },
\end{equation}
where $\delta = 10^{-4}$ is a small parameter \cite{BrownianMultiblobSuspensions} and
\[\V W^{f \tau}=\left[\V W_{1}^{f \tau}, \cdots, \V W_{N}^{f \tau} \right]^{T}, \ \V W^{D}=\left[\V W_{1}^{D}, \cdots, \V W_{N}^{D} \right]^{T}.\] 
Here random numbers are generated for each particle,
\[
\V W_{p}^{f \tau}= \left[ {a}^{-1}\V W_{p}^{f}, \V W_{p}^{\tau} \right],
\ \ \V W_{p}^{D}=\left[ {a} \V W_{p}^{f}, \V W_{p}^{\tau}\right], \ \ 1 \leq p \leq N
\]
where $\V W_{p}^{f}$, $\V W_{p}^{\tau}$ are $3 \times 1$ standard Gaussian random vectors. 

\begin{figure}
\begin{algorithm}
\singlespace
\caption{\label{alg:trap} \textbf{-- Stochastic Trapezoidal Split (STS) scheme} \\
For a given time step size $\D{t}$ and applied forcing $\V F\left(\V Q,t\right)$, this algorithm updates the configuration $\V Q^{n} \approx \V Q\left( t^{n} \right)$ at time $t^{n} = n \D{t}$ to $\V Q^{n+1}$. Orientations can be tracked using quaternions and updated by rotations, as described in \cite{BrownianMultiBlobs,BrownianMultiblobSuspensions}. Superscripts denote the time/configuration at which a quantity is evaluated, for example, $\V F^{n+1,\star} = \V{F}\left(\V Q^{n+1,\star},\,(n+1)\D{t}\right)$.} 
\singlespace
\small
\begin{enumerate}
\item \label{trap:0} Compute Brownian displacements (see section \ref{sec:BI})
\[
\D{\V{Q}}_W=\sqrt{\frac{2 k_{B}T}{\D{t}}} \Mob^{n} \left(\Lub^n\right)^{1/2} \M{W}_{1} + \sqrt{\frac{2 k_{B}T}{\D{t}}} \left(\Mob^n\right)^{1/2} \M{W}_{2}.
\]
\item \label{trap:1} Compute a predicted velocity $\V{U}^{n}$ by ignoring the drift term entirely and solving
\[
\left[\M{I} + \Mob^n \Lub^n  \right] \V{U}^{n} = \Mob^n \V F^{n} + \D{\V{Q}}_W,
\]
to give:
\begin{equation*}
\V{U}^{n} = \epsM^n \left( \V F^{n} + \sqrt{\frac{2 k_{B}T}{\D{t}}} \left(\Lub^n\right)^{1/2} \M{W}_{1} \right) + \sqrt{\frac{2 k_{B}T}{\D{t}}} \left[\M{I} + \Mob^n \Lub^n  \right]^{-1}  \left(\Mob^n\right)^{1/2} \M{W}_{2}. 
\end{equation*}
\item \label{trap:5}Compute the relevant RFD term $\V D^{\Mob}$ using \eqref{eq:RFD}, 
\begin{equation*}
\M{D}^{\Mob} = \frac{1}{\delta} \left[ \Mob\left(\V{Q}^n + \delta \V W^{D} \right)-\Mob\left(\V{Q}^n - \delta \V W^{D} \right) \right] \V W^{f \tau},
\end{equation*}
such that
\[
\av{\V D^{\Mob}} = \left(\partial_{\V Q} \cdot \Mob\right)^{n} + \sM O\left(\delta^{2}\right).
\]
\item \label{trap:6} Compute predicted configurations of the particles
\[
\V Q^{n+1,\star}=\V Q^{n}+\D{t} \, \V{U}^{n}.
\]
\item \label{trap:7} 
Compute corrected velocities by solving
\[
\left[\M{I} + \Mob^{n+1,\star} \Lub^{n+1,\star} \right] \V{U}^{n+1,\star}  = \Mob^{n+1,\star} \V F^{n+1,\star} +  \left( 2 k_{B}T\right) \V D^{\Mob} + \D{\V{Q}}_W,
\]
to obtain
\begin{align*}
\V{U}^{n+1,\star} &= \epsM^{n+1,\star} \V F^{n+1,\star}  + \left( 2 k_{B}T\right)  \left[\M{I} + \Mob^{n+1,\star} \Lub^{n+1,\star} \right]^{-1} \V D^{\Mob} \\
&+ \sqrt{\frac{2 k_{B}T}{\D{t}}}  \left[\M{I} + \Mob^{n+1,\star} \Lub^{n+1,\star} \right]^{-1}  \left(\Mob^n + \Mob^n \Lub^n \Mob^n \right)^{1/2} \V W_{1,2}.
\end{align*}

\item \label{trap:U} Update configurations to time $t+\D t$ using velocity $\V{U}^{n+1/2}=\left( \V{U}^n + \V{U}^{n+1,\star} \right)/2$,
\[
\V Q^{n+1}=\V Q^{n} + \D{t}\,\V{U}^{n+1/2}.
\]
\end{enumerate}
\end{algorithm}
\end{figure}

We show in appendix \ref{sec:proof} that step \ref{trap:U} of Algorithm \ref{alg:trap} indeed approximates equation \eqref{eq:Lang} with a weak accuracy of at least $\mathcal{O}\left(\D{t}\right)$. Specifically we show that the final configuration update in the STS scheme can be written as
\begin{align}
 \Delta \V Q^{n+1} &= \V Q^{n+1}-\V Q^{n} = \frac{\D{t}}{2} \left( \V U^{n} + \V{U}^{n+1,\star} \right) \\
 &= \frac{\D{t}}{2} \left( \epsM^n \V F^{n} + \epsM^{n+1,\star}  \V F^{n+1,\star} \right) + \sqrt{2 k_{B}T \D{t}} \, \left(\epsM^n\right)^{1/2} \V \, W_{1,2} \\
 &+ (k_{B}T) \D{t} \left( \partial_{\V Q} \cdot \epsM \right)^n  + \D{t} \, \mathcal{R}\left(\D{t},\D{t}^{1/2}\right),
\end{align}
where $\mathcal{R}\left(\D{t},\D{t}^{1/2}\right)$ denotes a Gaussian random error term with mean and variance of $\mathcal{O}\left(\D{t}\right)$. This trapezoidal update maintains second order accuracy in a deterministic setting ($k_B T = 0$), which helps improve the weak accuracy in the stochastic setting compared to the first-order Euler-Maruyama scheme used in \cite{SE_Multiblob_SD} (results not shown but see \cite{BrownianMultiblobSuspensions} for related studies).  We demonstrate the accuracy of our hydrodynamic model and the STS temporal integrator in Appendix \ref{app:MB_Stoch} by comparing to the rigid multiblob method \footnote{The rigid multiblob method we use here does not incorporate lubrication corrections but resolves the far-field hydrodynamics considerably more accurately than the RPY approximation.} previously developed by some of us in \cite{RigidMultiblobs,BrownianMultiblobSuspensions}.

\subsection{Firm repulsion between spheres} \label{firmpot}

Thermal fluctuations may introduce unphysical events such as particle--particle overlaps or particle--wall overlaps. For these unphysical configurations,  special care must be taken in defining the lubrication-corrected mobility so that overlaps occur rarely, and, should an overlap occur, the particles ought to separate quickly and through a thermodynamically reversible means.
Physically, there is a separation distance $\delta_{\text{cut}}$ below which additional physics enters (electrostatic repulsion, surface roughness, contact/friction forces, etc.). Motivated by this, we introduce a strong repulsive `firm' potential between particles and particles and the wall, and carefully modify $\epsM$ to accommodate the new contact dynamics.

The pairwise resistance resistance $\Rsuplub$ blows up when particles approach each other, and thus $\epsM$ will vanish. With a very small mobility, two nearly touching particles will tend to stay nearly touching unless acted upon by a large force. To push (nearly) overlapping surfaces apart when they are separated by less than $a \delta_{\text{cut}} $, we include a short-ranged but differentiable `firm' repulsive potential of the form \cite{MagneticRollers}
\begin{equation}\label{Usoft}
\Phi(r) = \Phi_0 
\begin{cases}
1 + \frac{d-r}{b_{\text{cut}}} & r < d \\
\text{exp}\left( \frac{d-r}{b_{\text{cut}}} \right) & r \geq d
\end{cases}.
\end{equation}
For particle-particle interactions, $r$ is the center-to-center distance and we take $d = 2a(1-\delta_{\text{cut}})$, and for particle-wall interactions $r$ is the particle center height and $d = a(1-\delta_{\text{cut}})$. We choose $b_{\text{cut}} = 2 a \delta_{\text{cut}} / \ln(10)$ as a cut-off length so that the inter-surface potential $\Phi \left(2a(1+\delta_{\text{cut}})\right) = 10^{-2} \Phi_0$. This ensures that the force is small when two surfaces are further than $a \delta_{\text{cut}} $ from touching and large when they overlap ($f_{\perp} = - \partial\Phi(r)/\partial r  \sim \Phi_0/b_{\text{cut}}$). We have found that taking $\delta_{\text{cut}} = 10^{-2}$  is sufficient for our purposes, and we use this value henceforth. 

The resistance correction $\Lub$ is not physically realistic for dimensionless surface separations (gaps) $\epsilon < \delta_{\text{cut}}$ ($\epsilon = r/(2a) - 1$ for pairs of particles, or $\epsilon = h/a - 1$ for a particle and a wall). A simple correction that we find to work fairly well is to take $\epsilon \leftarrow \max \left(\epsilon,\delta_{\text{cut}} \right)$.
This approach compliments the repulsive potential \eqref{Usoft}. Namely, the dimensionless perpendicular self--mobility coefficient of two overlapping surfaces is $X^{tt} \sim \delta_{\text{cut}}$, and therefore the relative radial separation velocity of two overlapping surfaces will be on the order of $u_{\perp} \sim X^{tt} f_{\perp} / (6 \pi \eta a) \sim \Phi_0/(\eta a^2)$. We use $\Phi_0 \sim 4 k_B T$ in this work to ensure that the repulsive energy for overlapping particles is larger than the thermal energy, but not so large as to require a sub-diffusive time step size. Thus, over a diffusive time scale $\tau_D \sim \eta a^3 / (k_B T)$, two overlapping surfaces will typically separate by a distance $\tau_D u_{\perp} \sim a$ on the order of the particle size, thus effectively eliminating the overlaps.

\section{Uniform Suspensions of Magnetic Rollers} \label{sec:MagRoll}

In past works, some of the authors have investigated active suspensions of rotating particles above a bottom wall, termed magnetic microrollers  \cite{Rollers_NaturePhys,MagneticRollers,NonlocalShocks_Rollers,BrownianMultiblobSuspensions}. The rotation of the particles is achieved in experiments by embedding a small cube of ferromagnetic hematite in each particle and applying a rotating magnetic field to the suspension \cite{Rollers_NaturePhys} (see the inset of figure \ref{fig:rollers} for a diagram of a typical roller suspension). The bottom wall couples the rotation of the particles to their linear velocity, and the coherence of the flow fields generated by each particle results in a greatly enhanced linear velocity for the whole suspension. 

In \cite{BrownianMultiblobSuspensions} a uniform suspension of rollers was simulated using the rigid multiblob method, and for sufficiently large packing densities ($\phi \sim 0.4$), a bimodal distribution was observed in the propulsion velocity of the particles. It was found that the bimodality of the velocity distribution is caused by a dynamic separation of the particles into two layers: a `slow lane' of particles whose center height was less than a particle diameter above the wall, and a `fast lane' of particle higher than one diameter above the wall. Previous experiments \cite{Rollers_NaturePhys} relied on PIV measurements of the suspension velocity, and could not capture a bimodal distribution. In this section we reinvestigate this problem using new particle-tracking-based experimental measurements, which do capture the bimodal distribution in the population velocity, and model the experiments using the more efficient numerical methods presented in this work. 

\subsection{Experimental Setup} \label{sec:Exp}

In our experiments, the suspensions of microrollers are composed of colloids with a ferromagnetic core suspended in water and driven by a rotating magnetic field. The spherical colloids are made of an off-center hematite cube embedded in 3-(trimethoxysilyl)propyl methacrylate (TPM) \cite{Sacanna2012}, which can be fluorescently labeled for imaging with fluorescence microscopy using 4-methylaminoethylmethacrylate-7-nitrobenzo-2-oxa-1,3-diazol (NBD-MAEM) \cite{Bosma2002,Youssef2019}.
The cubes have a side length of 770 nm (with 100 nm standard deviation) and have rounded edges.

We measured the size of the microrollers with both scanning electron microscopy (SEM) and dynamic light scattering (DLS), see Appendix~\ref{sec:exp_DLS} for details. From SEM, we found a diameter of 2.11$\pm$0.08~$\mu$m by measuring the diameter of 161 particles, which corresponds to a polydispersity (standard deviation/mean diameter) of 4\%. From DLS, we found a diameter of 2.03$\pm$0.04~$\mu$m. The particles were suspended in a 0.14 mM lithium chloride (LiCl) in water solution, which corresponds to a Debye length of $\sim$25 nm. We put the suspension in a glass sample cell with a height of $\sim$150 $\mu$m, as described in Appendix~\ref{sec:exp_cell}, and equilibrated for at least 30 minutes before imaging.

For the measurement of the diffusion constant $\bar{D}_{||}$ of the particles parallel to the floor, we imaged fluorescently labeled particles (see Appendix~\ref{sec:exp_diff}) at a very dilute concentration at a frame rate of 2 s$^{-1}$. The particle trajectories were determined using particle tracking \cite{Crocker1996, TrackPy}.

In order to determine the rolling velocity at different driving frequencies of dilute microrollers, we applied a rotating magnetic field using a home-built set of tri-axial nested Helmholtz coils \cite{Abbott2015}, placed on top of a fluorescence microscope as described in detail in Appendix~\ref{sec:exp_setup}. A magnetic field of 40 G, rotating around an axis parallel to the bottom glass wall, was applied and the fluorescently labeled particles were imaged at a rate of 9.0 s$^{-1}$. To prevent the particles from ending up at one side of the sample container, we inverted the direction of the rotating field every 30 seconds. We obtained the positions of the particles in the microscope images and linked them using particle tracking \cite{Crocker1996, TrackPy}, where overly bright (i.e. clusters) or stuck particles were left out of the analysis.

For the rolling experiments of dense suspensions, we mixed together particles with and without fluorescent labeling in a 1:1200 number ratio. This makes it possible to follow the dynamics of single rollers in a crowded layer using particle tracking. The area fraction of the monolayer of particles after sedimentation was estimated to be 0.4 by feature finding \cite{Crocker1996, TrackPy} in a single bright field microscope image, using the SEM estimate of the particle diameter.

\subsection{Simulation parameters}\label{sec:params}

In order to determine appropriate parameters for the simulations, we use a very dilute suspension to experimentally measure key parameters for an isolated microroller. The diffusion constant $\bar{D}_{||}$ of an isolated particle parallel to the glass wall was measured to be $\bar{D}_{||}$ = 0.103$\pm$0.003 $\mu$m$^2$s$^{-1}$, from a total of 21,000 displacements.

\begin{figure}
\centering
\includegraphics[width=0.75\textwidth]{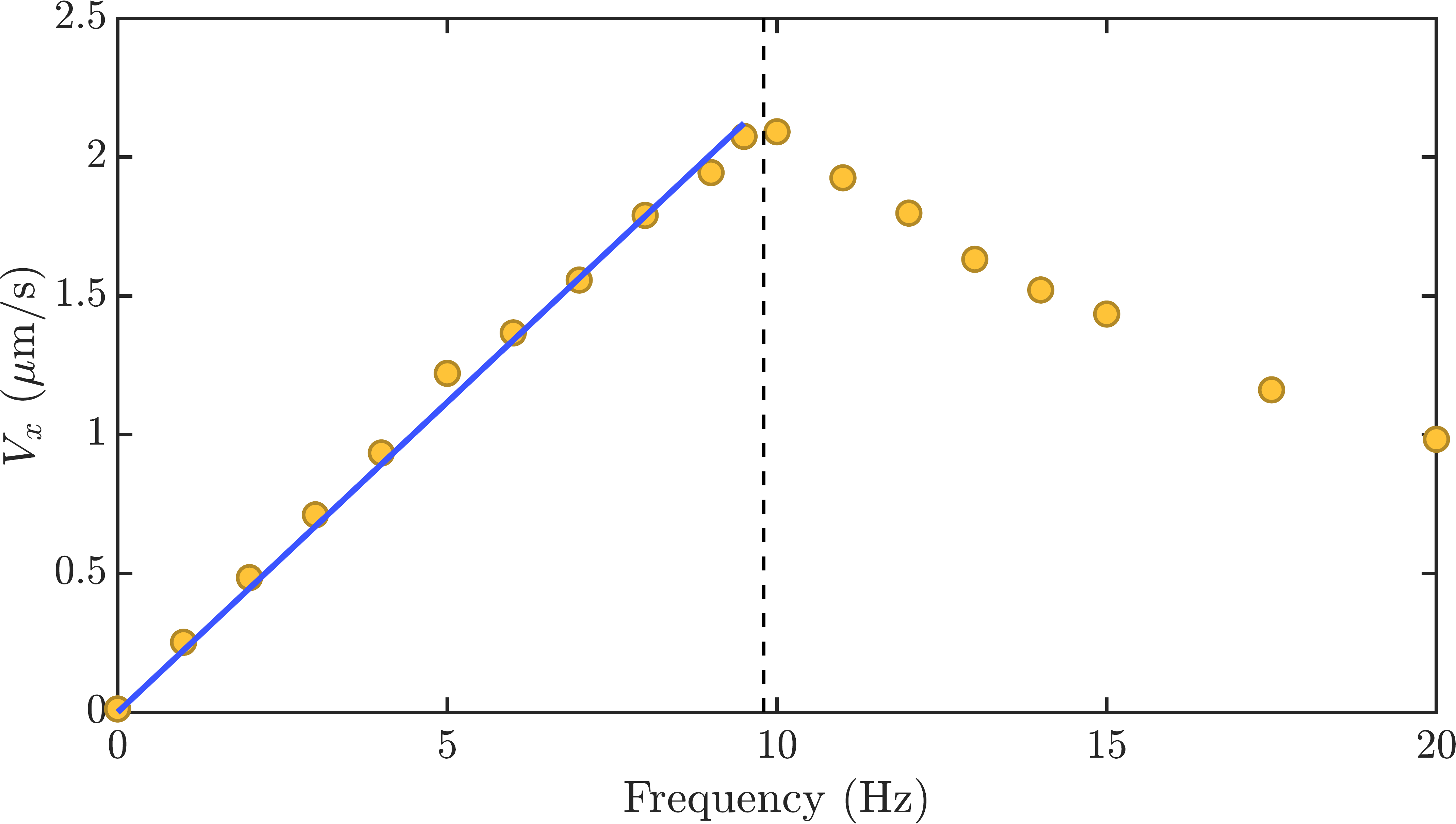}
\caption{Measured velocity of dilute microrollers as a function of the frequency of the applied rotating magnetic field (40 G). Up to a frequency of $\sim$9.8 Hz (vertical black dashed line), the velocity increases linearly with the frequency (blue line, $\text{slope}=\bar{A}_f = 0.223~\mu$m). At higher frequencies the velocity decreases for increasing frequency, as the rollers cannot overcome the viscous torque of the surrounding liquid.}
\label{fig:v_vs_f}
\end{figure}

We also measured the average velocity of dilute fluorescent rollers driven by a 40 G magnetic field for frequencies up to 20 Hz, see Fig.~\ref{fig:v_vs_f}. Up to a frequency of $\sim$9.8 Hz (black dashed line), the velocity of the rollers increases linearly with the frequency of the applied rotating magnetic field with a slope of $\bar{A}_f = 0.22 \ \mu$m. Above this frequency the velocity starts to decrease upon an increase in the frequency. This is due to the inability of the particles to overcome the viscous torque of the surrounding liquid as the particles start to slip relative to the field \cite{Rollers_NaturePhys}. To prevent this, we use a frequency of 9 Hz in our dense suspension experiments, and confirm using simulations that the slippage is minimal.

The ambient room temperature for the experiments was $T = 22~^{\circ}$C, and therefore the viscosity of water is taken to be $\eta = 0.96 \times 10^{-3}$ cP. We use the DLS measurement of the particles' radius and take $a = 1.02~\mu$m.
Using SEM measurements, the volume of hematite core of the particles was estimated to be $V_{\text{core}} \approx 0.95 \times (770~\text{nm})^3$ (where the $0.95$ factor corresponds to a $5\%$ loss in volume from rounded edges). 
Using literature values for the density of hematite and the TPM colloid \cite{Hematite_mass,TPM_mass}, we estimate the buoyant mass $m_e$ of the particles as $3.1 \times 10^{-15}$ kg. 

The equilibrium Gibbs--Boltzmannn distribution for the height $h$ of a single particle sedimented above the bottom wall is 
\begin{align}
P_{GB}(h) \propto \exp \left( -\left(m_e g h + U(h) \right)/ k_B T \right). \label{eq:GB}
\end{align}
The steric potential $U(h)$ is
$U(h) = U_{\text{firm}}(h) + U_{\text{soft}}(h)$,
where $U_{\text{firm}}$ is the firm potential described in section \ref{firmpot}, and $U_{\text{soft}}$ is a soft potential of the form \eqref{Usoft} which captures the electrostatic repulsion from the bottom wall. We also include a soft, pairwise repulsion between particles with the same form as $U_{\text{soft}}$. 

The excess mass $m_e$, the strength of the soft potential $\Phi_{\text{s}}$, and the effective Debye length $b_{\text{s}}$ are difficult to measure precisely, and combine together to control the typical height of the particles above the wall. To estimate suitable values of $\Phi_{\text{s}}$ and $b_{\text{s}}$ for our simulations, we fix $m_e = 3.1 \times 10^{-15}$ kg, and try to match the experimentally measured values of the parallel diffusion coefficient $\bar{D}^{||}$, and the slope of $V(f)$ for $f<f_c$, $\bar{A}_f$, described in section \ref{sec:Exp}. We compare these measurements to numerical estimates computed by averaging the lubrication-corrected mobility for an isolated particle 
over the equilibrium Gibbs--Boltzmann distribution \eqref{eq:GB},
\begin{align}
 D^{||} &= k_B T \av{\hat{\V{x}}^{T} \epsM^{tt} \hat{\V{x}}}_{GB}, \\
 A_f &= V'(f<f_c) = 2 \pi \av{\hat{\V{x}}^{T} \epsM^{tr} \left( \epsM^{rr} \right)^{-1} \hat{\V{y}}}_{GB}. 
\end{align} 
We numerically find the values of $\left( \Phi_{\text{s}},b_{\text{s}} \right)$ which minimize the total relative error with experiments
\begin{equation}
 \text{Error} = \sqrt{ \left( \frac{D^{||}-\bar{D}^{||}}{\bar{D}^{||}} \right)^{2} + \left( \frac{A_f-\bar{A}_f}{\bar{A}_f} \right)^{2} }. \label{eq:errordpar}
\end{equation}
While this error never completely vanishes, we find that taking $\Phi_{\text{s}} \approx 8 k_B T$  and $b_{\text{s}} \approx 0.04 a \approx 40$ nm minimizes the error at about $11.5 \%$, and we use those values in the rest of this section.  Note that the selected value of $b_{\text{s}}$ is consistent with the $\sim$25 nm  Debye length estimated from the experimental parameters.

Figure \ref{fig:v_vs_f} in section \ref{sec:Exp} shows that a single particle begins to `slip' behind the magnetic field when the angular velocity of the field $\Omega > 2 \pi (f_c=9.8 \text{Hz}) = ||\V \omega_{c}||$. The constant torque $ \V \tau_{c}$ required to rotate an isolated particle with an average angular velocity of $\V \omega_{c} = \Omega_c \hat{\V{y}}$ satisfies
\[
 \av{\epsM^{rr}}_{GB} \V \tau_{c} = \V \omega_{c},
\]
and we compute $||\V \tau_{c}|| = 2.0 \times 10^{-18}$Nm. This is the maximal torque $\V \tau = \V m \times \V B$ that the magnetic field can exert on any particle, where $\V{m}$ is the magnetic moment of the hematite. From $\tau_{c} = m B$ we compute the strength of the magnetic moment in the particles as $m=||\V m || = 5.0 \times 10^{-16} \text{Am}^{2}$ (using $B = 40$G), in perfect agreement with the estimate given in \cite{Rollers_NaturePhys}.

\subsection{Dense suspensions} \label{sec:Sims}

\begin{figure}
\includegraphics[width=1\textwidth]{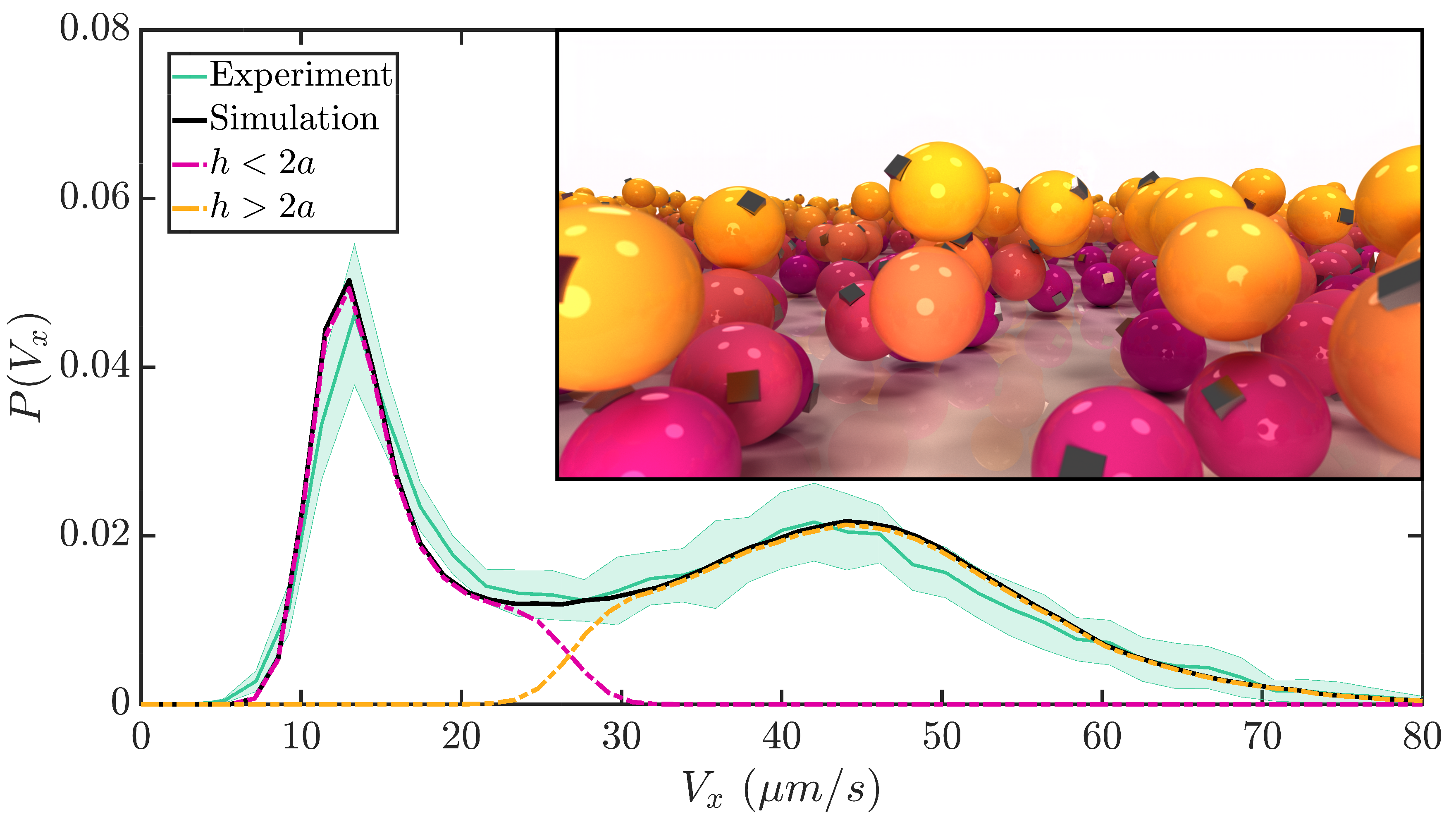}
\caption{The distribution of velocities in the direction of collective motion for the microroller suspension shown in the inset. We compare the experimentally measured distribution (green solid line) with the distribution computed using our lubrication-corrected Brownian dynamics method (black solid line). The experimental data represents the mean over 8 independent runs, and the extents of the shaded area represent $95\%$ confidence bounds.  The simulated data is broken into two sub-distributions according to the height of the particles above the wall ($h<2a$ or $h>2a$), showing a clear correlation between the `slow' peak in the velocity distribution and the lowest particles (with a similar correlation for fast and high particles).
Inset: A typical configuration for a uniform suspension of microrollers at density $\phi=0.4$ and driving frequency $f=9$Hz. The hematite cube embedded in the particles is overemphasized here for visual clarity. Low (slow) particles are colored magenta while high (fast) particles are colored yellow. \modified{Also see supplementary media for videos from simulations as well as experiments.}}
\label{fig:rollers}
\end{figure}

We experimentally measured the trajectories of the microrollers in a dense suspension (in-plane packing fraction $\phi \approx 0.4$) in a rotating magnetic field (40 G, 9 Hz).
The effective (apparent) particle velocities in the direction of bulk motion ($x$-direction) were computed over a time interval of $1$ s. Fig.~\ref{fig:rollers} shows the probability distribution of particle velocities $P(V_x)$. The histogram was computed by averaging eight independent $30s$ runs and the shaded region around the `Experiment' curve shows the $95\%$ (2 std.) confidence bounds. Also included in Fig.~\ref{fig:rollers} is an analogous velocity distribution computed from simulations of this uniform roller suspension, described next. The agreement between the simulated and measured bimodal distributions is quite good, and demonstrates that the lubrication-corrected BD method has quantitative accuracy sufficient to reproduce the experimental measurements.   

Figure \ref{fig:rollers} also shows sub-distributions of the simulated $P(V_x)$ wherein the particle velocities are grouped into high particles (whose height $h > 2a$ from the bottom wall) and low particles ($h <2a$). While there is some small overlap, it is quite clear that the low particles correspond to the slow peak in $P(V_x)$, and the high particles correspond to the fast peak, as originally observed in \cite{BrownianMultiblobSuspensions}. For the first time, we show here that the peaks of the sub-distributions corresponding to $h > 2a$ and $h < 2a$ closely coincide with the peaks of the experimentally measured bimodal distribution.     

\subsubsection{Simulations of dense uniform suspensions}

Figure \ref{fig:rollers} shows results for the distribution of propulsion velocities obtained by simulating a uniform suspension with a packing density $\phi \approx \pi a^2 N/L^2 = 0.4$, where $N$ is the number of particles in the square domain and $L$ is the length of the domain.
We use $N=2048$ particles and periodic boundary conditions (implemented using periodic images as in \cite{MagneticRollers}). We confirmed that the number of particles is large enough that periodic artifacts are negligible by computing the velocity distribution for a larger domain size that include one periodic image in each direction, i.e., $N = 9\times 2048$ particles.    

Following our experiments we compute (apparent) particle velocities over intervals of one second for all of the distributions presented in this section. By convention we take the direction of applied magnetic field to be in the $\hat{\V{y}}$ direction and compute statistics of the particles' velocity $V_x$ in the $x$-direction. Velocity distributions are computed as a normalized histogram of the apparent velocities using 1500 samples taken after a sufficiently long period of equilibration.

\begin{figure}
\centering
\begin{subfigure}{.85\textwidth}
  \centering
  \includegraphics[width=\textwidth]{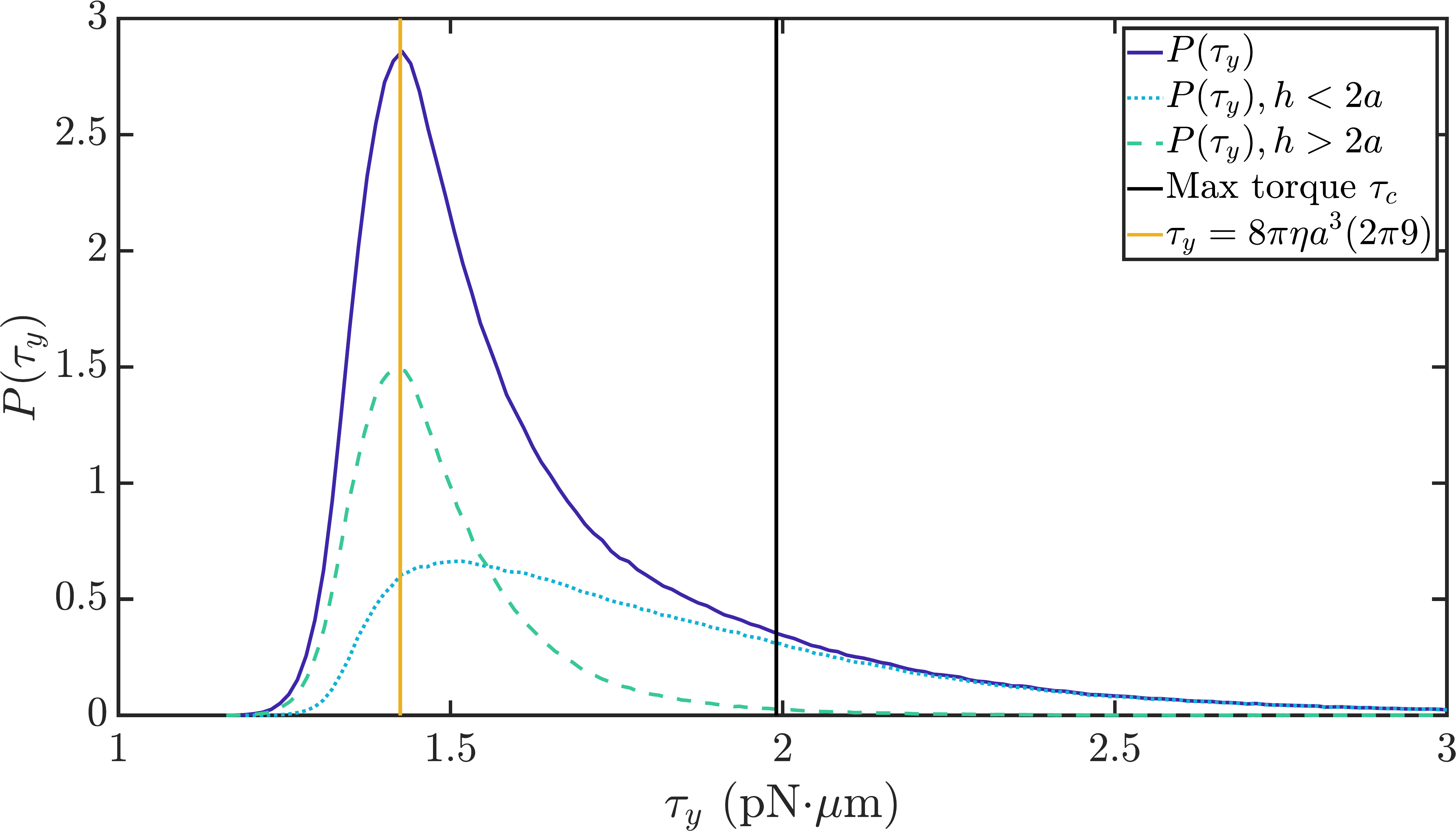}
  \caption{Panel A}
\end{subfigure}\\
\begin{subfigure}{.85\textwidth}
  \centering
  \includegraphics[width=\textwidth]{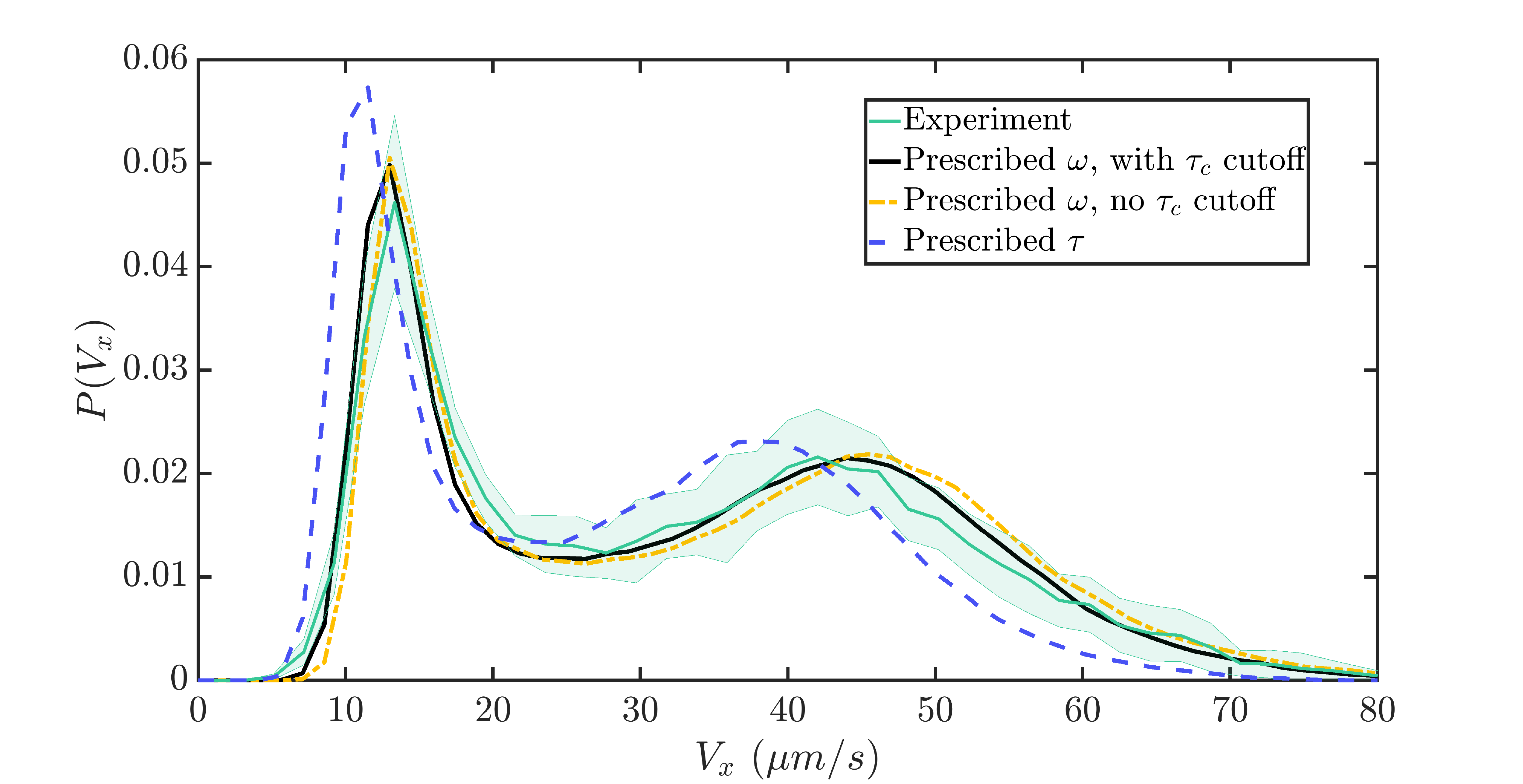}
  \caption{Panel B}
\end{subfigure}
\caption{\label{fig:TcO}(Panel A) Probability distribution of applied torques required to maintain an approximately constant angular velocity $\V \omega$ for all particles. The distribution is grouped into particles whose center is above $2a$ from the wall and those below, which Fig. \ref{fig:rollers} shows correspond to fast and slow particles, respectively. The low (slow) particles dominate the tail of the torque distribution. Also shown is a yellow line representing the constant torque approximation $ \tau_y = 8 \pi \eta a^3 \omega$. The solid black line represents the `slip' limit where the applied torque exceeds $\tau_c = m B$. (Panel B) Comparison of velocity distributions $P(V_x)$ when the particles are driven either by a prescribed angular velocity $\omega \hat{\V{y}}$, with and without a cutoff of $\tau_c$ for the applied torque, or a prescribed torque $\V \tau = 8 \pi \eta a^3 \omega \hat{\V{y}}$.}
\end{figure}

Figure \ref{fig:v_vs_f} confirms that magnetic rollers driven by an AC magnetic field below the critical frequency ($f_c = 9.8$ Hz) rotate coherently with the magnetic field. Following section \ref{sec:constrain}, we compute the applied torques $\V \tau_{\omega}$ required to constrain the angular velocity of each particle to be $\V \omega = 2 \pi (9 \text{Hz}) \hat{\V{y}} = \Omega \hat{\V{y}}$ in the absence of Brownian motion.  Panel A of Fig. \ref{fig:TcO} shows the distribution of torque magnitudes $||\V \tau_{\omega}|| = \tau_{\omega} \approx \left[\V \tau_{\omega}\right]_y$, with a black vertical bar demarcating the slip cutoff $\tau_{\omega}=\tau_{c}$. We see that the torques are broadly distributed with a long tail including torques larger than $\tau_c$, dominated by slow particles with $h<2a$. In panel A of Fig. \ref{fig:TcO} we also show that a constant torque with $||\V \tau|| = 8 \pi \eta a^3 \omega$ (as was used in \cite{BrownianMultiblobSuspensions}) correctly estimates the most probable torque, without, however, accounting for the broad distribution of torques.

To account for the upper bound $\tau_c= m B$ on the magnitude of the torque exerted by the applied field, we cap the applied torque and define
\[
\widetilde{\V \tau}_{\omega} = \frac{\min \left(  \tau_{c} ,  \tau_{\omega}\right)}{\tau_{\omega}} \V \tau_{\omega}.
\] 
Panel B in Fig. \ref{fig:TcO} shows velocity distributions from suspensions driven by applying a torque $\widetilde{\V \tau}_{\omega}$ (solid black line, also included in Fig. \ref{fig:rollers}) or $\V \tau_{\omega}$ (dashed-dotted orange line). The difference between using $\widetilde{\V \tau}_{\omega}$ over $\V \tau_{\omega}$ is small compared to the experimental and statistical uncertainties. Panel B also shows $P(V_x)$ for a suspension driven by applying $\V \tau = 8 \pi \eta a^3 \Omega \hat{\V{y}}$ (dashed blue line), which clearly maintains the qualitative features of the experimental velocity distribution (e.g. bimodality, and relative mass of the modes), but provides a notably worse quantitative agreement with our experiments. In Appendix \ref{app:MB_Stoch} we compare the propulsion velocities computed using the lubrication-corrected BD method (for constant applied torques) to reference results computed using the rigid multiblob method \cite{BrownianMultiblobSuspensions}. We find a very good agreement with the results obtained using 42 blobs per colloid, which is considerably more expensive than our minimally-resolved approach that uses one blob per colloid for the far-field hydrodynamics.

  
\subsubsection{Switching Lanes}

Figure \ref{fig:rollers} shows that we can separate the two peaks in the velocity distribution of the roller suspension by the height of the colloids. The fast peak roughly corresponds to particles whose center $h$ is above a distance of $2a$ from the wall and the slow peak to particles below $2a$. These lanes form as a result of the driven dynamics in the suspension, and it is natural to ask how often a particle changes lanes.

\begin{figure}
\centering
\begin{subfigure}{.7\textwidth}
  \centering
  \includegraphics[width=\textwidth]{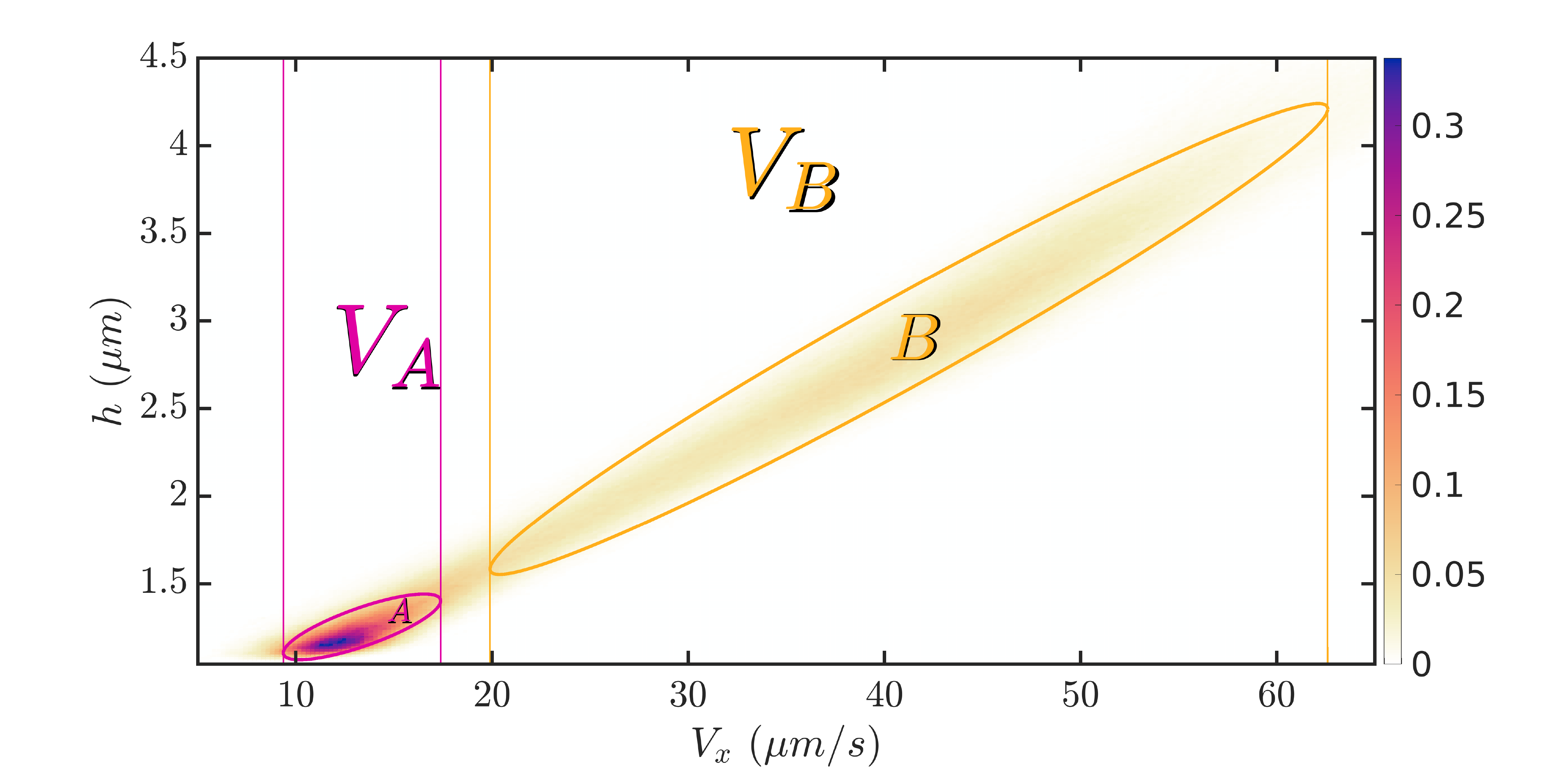}
  \caption{Panel A}
  \label{fig:rollersub1}
\end{subfigure}\\
\begin{subfigure}{.75\textwidth}
  \centering
  \includegraphics[width=\textwidth]{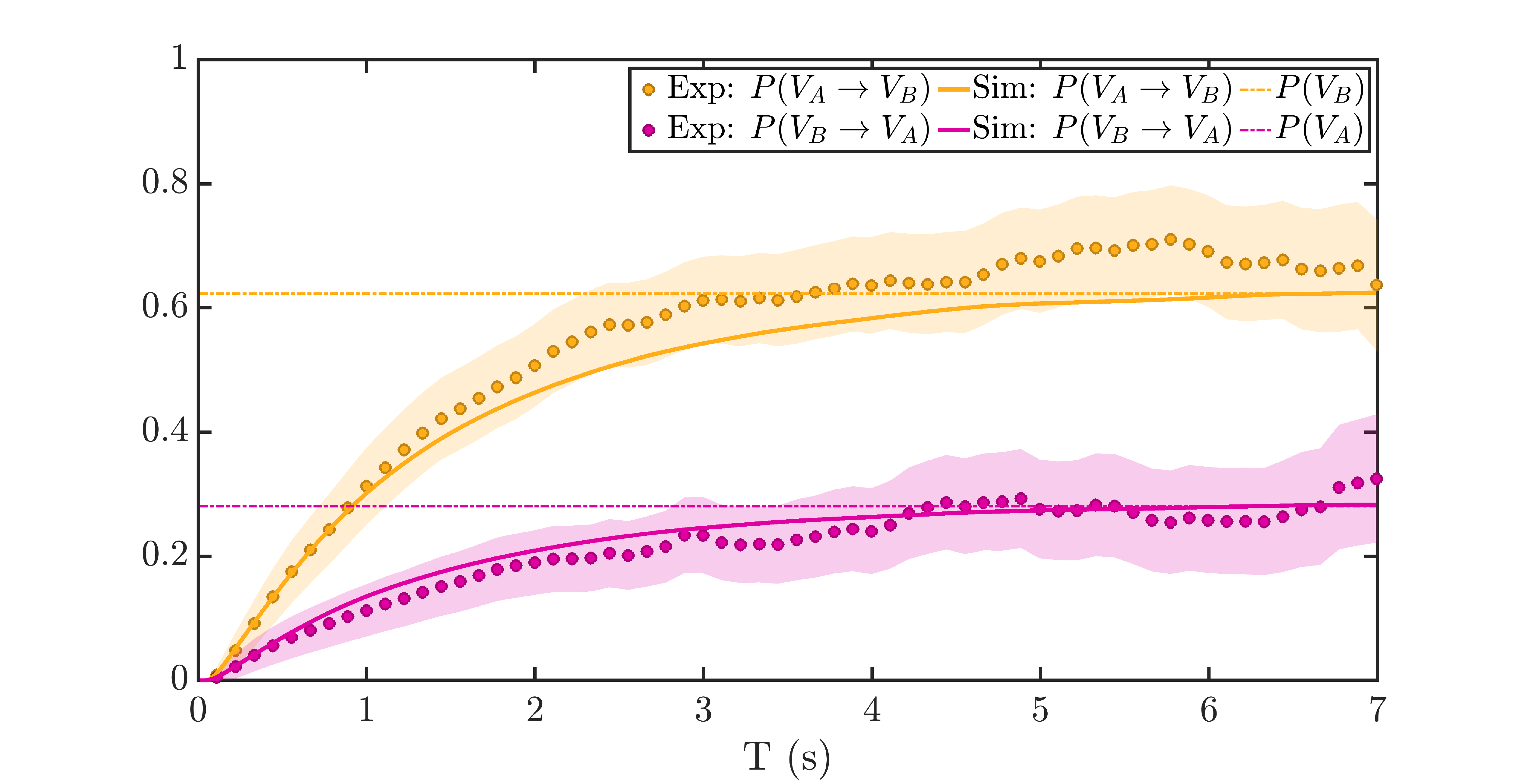}
  \caption{Panel B}
  \label{fig:rollersub2}
\end{subfigure}
\caption{\label{fig:P_AB}(Panel A) Pseudocolor map of the joint steady-state distribution $P(V_x,h)$ of particle velocities and heights, computed from the simulation data. Two elliptical regions demarcate regions we have identified as the `fast-lane' (region B) and the `slow-lane' (region A). The one dimensional intervals $V_A$ and $V_B$ demarcated by color-coded vertical lines correspond to the $V_x$ extents of the sets $A$ and $B$, respectively. (Panel B) The probability of a particle starting in set $V_{A/B}$ to end up in set $V_{B/A}$ after a time $T$. Simulated data is shown as solid lines which asymptote to $P(A)$ or $P(B)$ depending on the state that the particle's trajectory is conditioned to arrive in. Experimental data, shown as circular markers, agrees with our simulated data within a $95\%$ confidence interval ($2$ std.),  shown as a shaded region.}
\end{figure}

Using our simulation data, we can compute the joint distribution function $P(V_x,h)$ for the particles' height and velocity. Panel A in Fig. \ref{fig:P_AB} shows a pseudocolor map of $P(V_x,h)$, where we identify two elliptical regions corresponding to the modes (peaks) of the distribution, readily identified as the slow (region $A$) and fast (region $B$) lanes. The elliptical regions are identified by fitting a bimodal Gaussian mixture model to $P(V_x,h)$, and we have plotted level sets corresponding $95\%$ of the probability mass in each mode, separately. The large eccentricity of these elliptical regions quantifies our observation that height and velocity in the suspension are highly correlated. Hence to identify which lane a particle resides in we only look at its velocity, allowing us to compare simulated results with experimental ones. Specifically, we use the velocity extrema of groups $A$ and $B$ to define the intervals $V_A=[9.37,17.4] \mu\text{m/s}$ and $V_B=[19.9,62.6]\mu\text{m/s}$ respectively as the `slow' and `fast' lanes (shown in Panel A of Fig. \ref{fig:P_AB} as color coded vertical lines). The probability of a particle occupying these groups is calculated as $P(V_A) = 0.28$ and $P(V_B) = 0.62$.

To interrogate how often a particle will switch lanes, say from the slow to the fast lane, we compute the probability $P(V_A \rightarrow V_B)$ that a particle will be in $V_B$ at time $t=T$, given that it started in $V_A$ at $t=0$. At long times, a particle will forget where it started and $P(V_A \rightarrow V_B)$ will asymptotically approach $P(V_B)$, as seen in Panel B of Fig. \ref{fig:P_AB}. To compute an unbiased estimator for $P(V_A \rightarrow V_B)$, we consider segments of particles' trajectories which start in $V_A$ at $t=0$ or enter $V_A$ at a certain time $t$, and check whether they end up in $V_B$ a time $T$ later. The variance of the estimate for $P(V_A \rightarrow V_B)$ at each time $T$ can be computed as the variance of the average of $N_t$ independent binomial variables, $\text{var}\left(P\right) \approx P(1-P)/N_t$, since the $N_t$ trajectory snippets (samples) are approximately statistically independent.   

The switching dynamics can be modeled as a simple two-state Markov model for the lane changing dynamics where a particle will switch from $V_A$ to $V_B$ with rate $r_{AB}$ and vice versa with rate $r_{BA}$, giving
\[
\frac{P(V_A \rightarrow V_B) }{P\left( V_B \right)}  = 
\frac{P(V_B \rightarrow V_A) }{ P\left( V_A \right)} =
1 - \exp \left(- \frac{t}{\tau_{AB}}  \right) \label{eq:PAB1}
\]
where $\tau_{AB} = P\left( V_B \right)/r_{AB} = P\left( V_A \right)/r_{BA}$. These predictions match the simulation data for $\tau_{AB}=1.5$s ($r_{AB} = 0.42$ and $r_{BA} = 0.19$).

Panel B of Fig.~\ref{fig:P_AB} compares experimentally measured values of $P(V_A \rightarrow V_B)$ and $P(V_B \rightarrow V_A)$ against simulations. Note that the particle trajectories measured in our experiments range in duration from $3$ s to $25$ s and are therefore not long enough to accurately sample the long-time behavior. Nevertheless, we see good agreement in the switching dynamics between experiments and simulations, showing again that our lubrication-corrected BD method models the driven dynamics with quantitative accuracy.

\section{\label{sec:Sheet}Lubrication friction in a dense monolayer of microrollers}

In \cite{MIPS_Quincke_Bartolo}, Geyer \emph{et al.} showed experimentally that a suspension of Quincke rollers can self separate into a dense active solid phase and a sparse `polar' phase. By increasing the average packing density of the system, they observe that the average velocity of the suspension initially increases with density but eventually becomes an `active solid' where the velocity of the suspension is retarded to the point of arrest. In appendix A of \cite{MIPS_Quincke_Bartolo}, the authors conjecture that this dynamic arrest seen in their experiments is due to inter-particle lubrication interactions frustrating the motion of the suspension at high in-plane packing fractions. Specifically, they conjecture that the arrest happens when there is a balance between viscous torque from inter-particle lubrication and the applied electrodynamic torque. In this section, we interrogate whether lubrication interactions cause a dynamic arrest in dense suspensions of microrollers driven by a constant applied torque, rather than attempting to simulate the complex electrohydrodynamics of Quincke rollers \cite{MIPS_Quincke_Bartolo}. 

In the following simulations, we take the particle radius $a=1~\mu$m. As in section \ref{sec:MagRoll}, we will take $\eta = 0.96 \times 10^{-3}$ cP and $T = 22~^{\circ}$C. We confine the particles to remain approximately fixed in a plane above the bottom wall at a height $h_c = 1.1 a$ using a strong harmonic potential
\[
\Phi_{c}(h) = 10^3 k_B T \left(h - h_c \right)^{2},
\]
and therefore we neglect gravity. The height $h_c$ is taken to be very close to the wall to mimic the experiments of \cite{MIPS_Quincke_Bartolo}. The strength of the potential was chosen through numerical experimentation to ensure that the particles remain strictly fixed in the desired plane $h=h_c$ even at high packing densities. Following what was done in section \ref{sec:MagRoll}, we include a soft repulsive potential between the particles in the form of \eqref{Usoft} with $b_{\text{cut}} = 0.1 a$ and $\Phi_0 = 4 k_b T$, in \textit{addition} to the `firm potential' discussed in section \ref{firmpot}. 

We take the geometry of our domain to be semi-infinite in $z$ and periodic in $x$ and $y$, and use a fixed number of particles $N=1024$ in every simulation. We use the periodic domain size $L$ to control the in-plane packing fraction $\phi = \pi a^2 N/L^2$. To interrogate the dependence of velocity on packing fraction, we examine suspensions driven by a constant torque
\[
 \V{T}_{c} = 8 \pi \eta a^3 \Omega_{c} \hat{\V{y}} = 8 \pi \eta a^3 \left( 2 \pi (1 \text{Hz}) \right) \hat{\V{y}},
\]
as well as suspensions driven to maintain a constant rotation rate $\V \omega_c = \Omega_{c} \hat{\V{y}}$ using the method described in section \ref{sec:constrain}. The rolling motion of each particle generates a net translation in the $x$ direction with a steady state velocity distribution $P\left(V_x\right)$. We take the velocity of the whole sheet to be the mean of this distribution $V \equiv \av{V_x}$, and study the dependence $V\left( \phi \right)$. 

\begin{figure}
\centering
\begin{subfigure}{.45\textwidth}
  \centering
  \includegraphics[width=\textwidth]{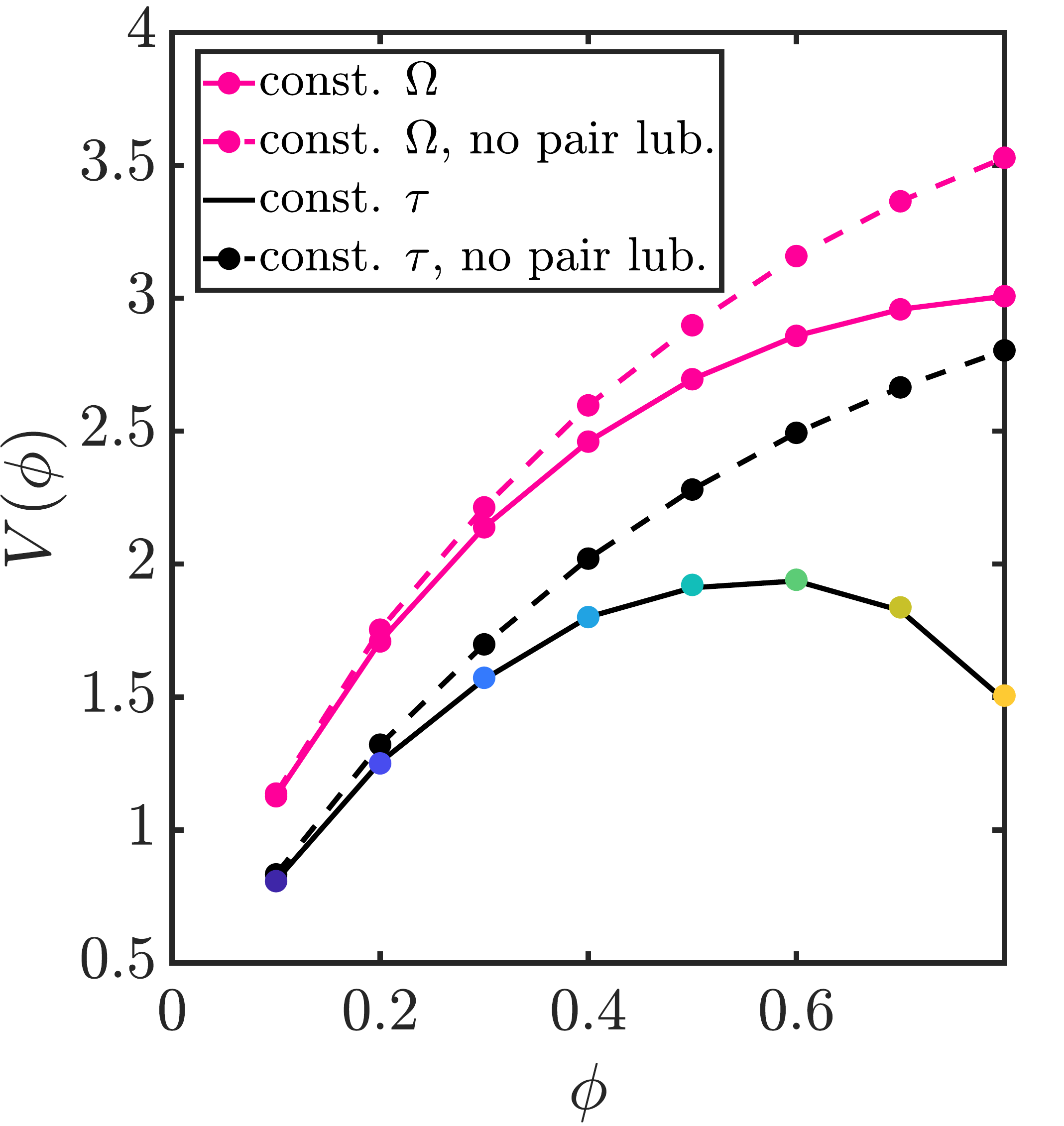}
  \caption{Panel A}
\end{subfigure}
~
\begin{subfigure}{.45\textwidth}
  \centering
  \includegraphics[width=\textwidth]{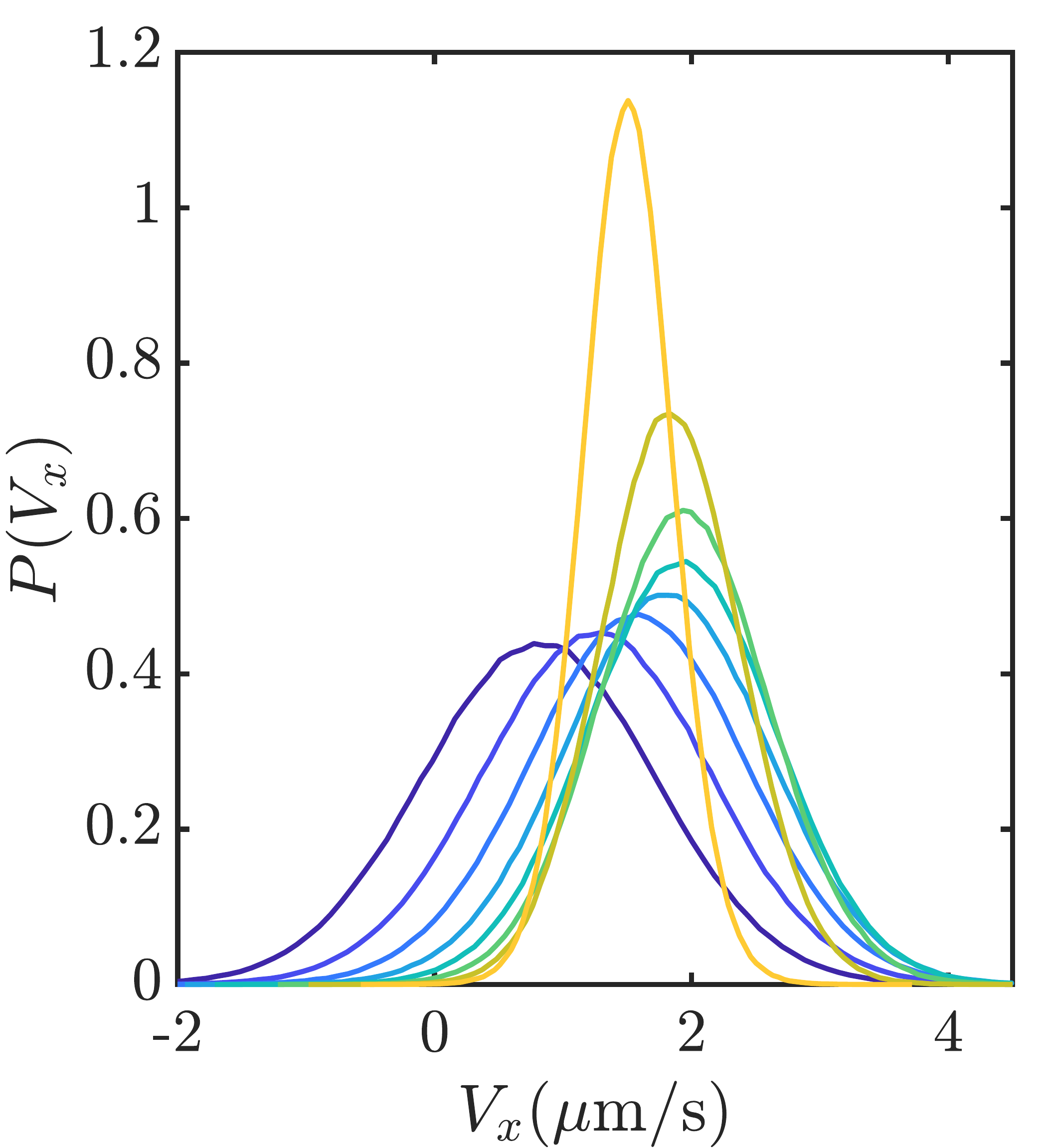}
  \caption{Panel B}
\end{subfigure}
\caption{\label{fig:Vphisheet} (Panel A) Comparison of two driving mechanisms for a suspension of rollers which are confined by a strong harmonic potential to remain fixed in a plane separated by a gap of $0.1a$ from the bottom wall. \modified{Each curve} shows the mean velocity \modified{(in $\mu \text{m/s}$)} in the direction of collective motion ($x$-direction) as a function of the in-plane packing fraction. The solid black curve corresponds to particles driven by a constant applied torque $\V \tau = 8 \pi \eta a^3 (2 \pi) \hat{\V{y}}$, and the solid fuchsia curve corresponds to particles driven by constraining their angular velocity to a constant value $\V \omega = 2 \pi \hat{\V{y}}$. When the particles are driven by a constant torque there is a clear peak in $V(\phi)$ around $\phi \approx 0.5$. On the other hand, particles driven by a constant $\Omega$ show a steady, though diminishing, increase in velocity as $\phi$ is increased. Also shown as dashed lines are $V(\phi)$ profiles for both driving mechanisms but \textit{without} pairwise lubrication forces between neighboring particles. (Panel B) Velocity distributions in the $x$-direction for constant applied torque with pairwise lubrication friction. \modified{The color of each curve corresponds to the packing density $\phi$ used in the simulation, coded according to the corresponding colored marker on the solid black curve in Panel A. Also see supplementary media for a video of one of the simulations.}}
\end{figure}

Figure \ref{fig:Vphisheet} shows a bulk slowdown of the suspension at high densities, $\phi > 0.5$, when a constant torque is applied (the solid black curve with colored markers). Panel B in this figure shows that the velocity distribution $P(V_x)$ for each packing density is approximately Gaussian, with a variance that narrows as $\phi$ is increased. Hence, as $\phi$ is increased the particles tend to move with a more uniform velocity as would a `solid' phase. Both the maximum in the plot of $V(\phi)$ vs $\phi$, and the narrowing variance in $P(V_x)$ as $\phi$ is increased, are due to the increasing lubrication force between nearly touching particles as $\phi$ is increases. To show that the lubrication between particles retards their hydrodynamic responsiveness to applied torques, we turn off the pairwise lubrication corrections \footnote{The lubrication corrections with the bottom wall are still included.}. The dashed black curve in Fig. \ref{fig:Vphisheet} shows that when pairwise lubrication corrections are \textit{not} included, $V(\phi)$ exhibits a monotonically increasing dependence on $\phi$.      
 
If we change the driving mechanism of the particles in the sheet from a constant applied torque to a prescribed angular velocity $\V{\omega} = \Omega_c \hat{\V{y}}$, we see a marked change of behavior in the $V(\phi)$ curve. The solid teal line in Fig. \ref{fig:Vphisheet} shows a monotonic growth in $V(\phi)$ as $\phi$ is increased. This is not so surprising. One needs to generate a large enough applied torque so that $\V{\omega}$ remains constant regardless of packing density, thus overcoming the lubrication force. In practice, however, the maximum torque must be limited by the physical driving mechanism, for both Quincke rollers and magnetic particles. When we remove the effects of pairwise lubrication for particles driven by a constant angular velocity, the trend in $V(\phi)$ (the dashed teal line in Fig. \ref{fig:Vphisheet}) is similar to when pairwise lubrication was included but with a less pronounced saturation in the growth of $V(\phi)$ for larger $\phi$.

\section{Conclusions}

We reported new experimental and computational results on the collective dynamics of a dense suspension of colloids sedimented above a bottom wall and spun by a rotating magnetic field. The experiments used fluorescent tracers to enable precise measurements of the motion of individual active particles. We also developed a lubrication-corrected Brownian Dynamics method for driven suspensions of spherical colloids confined above a bottom wall. We showed that our numerical method can predict both static and dynamic nonequilibrium statistics of a driven Brownian suspension of microrollers accurately enough to provide \textit{quantitative} agreement with our experiments. Specifically, both simulations and experiments showed a bimodal distribution of the particles' velocities, with good agreement about the locations and widths of the two peaks. The two sub-populations of microrollers correspond to particles in a slow layer right above the floor, and a faster layer above the first layer. We showed good agreement between simulations and experiments on the distribution of switching times between the two sub-populations of particles. The accuracy of our minimally-resolved simulation method is owed, in no small part, to the improved hydrodynamic accuracy provided by lubrication corrections for pairs of nearby surfaces (particles and the bottom wall or pairs of particles).

We also showed numerically that lubrication forces between nearly touching particles in a dense suspensions of rollers are a plausible explanation for the formation of the active solid phase observed in \cite{MIPS_Quincke_Bartolo}. Our suspension of microrollers does not exhibit a sharp motility-induced phase separation (MIPS), at least for the system sizes studied here. However, the collective slowdown for $\phi > 0.5$ is qualitatively similar to that seen for Quincke rollers in \cite{MIPS_Quincke_Bartolo}. Specifically, we saw that when a constant torque is used to drive particles in the suspension, the average velocity of the suspension $V$ has a maximum at packing density $\phi \approx 0.5$, and that this maximum is directly caused by pairwise lubrication between particles. The stark difference in trends in $V(\phi)$ when angular velocity or torque is prescribed agrees with the results that we saw for magnetic rollers. Together, these examples demonstrate that the collective dynamics of dense suspensions held close to a bottom wall, for which lubrication plays a big role, is strongly affected by the driving mechanism.

\begin{acknowledgments}
We thank Paul Chaikin and Blaise Delmotte for numerous informative discussions regarding microrollers\modified{, and an anonymous reviwer for bringing work of Cichocki and Jones to our attention}. B.S. and A.D. thank Adam Townsend, Helen Wilson, and James Swan for their help with constructing accurate pairwise lubrication formulas, and Eric Vanden-Eijnden for informative discussions about analyzing bistable dynamics. E.B.v.d.W. and M.D. thank Mena Youssef and Stefano Sacanna for providing the hematite-TPM particles and Kevin Ryan and Venkat Chandrasekhar for help with 3D printing the frame for the Helmholtz coil set.
This work was supported by the National Science Foundation under award number CBET-1706562. B.S. and A.D. were supported by the National Science Foundation via the Research Training Group in Modeling and Simulation under award RTG/DMS-1646339, by the MRSEC Program under award DMR-1420073. B.S. and A.D. also thank the NVIDIA Academic Partnership program for providing GPU hardware for performing the simulations reported here. 
\end{acknowledgments}

\appendix

\section{\label{apdx:RPRW}Semi-Analytical formulas for the pair resistance matrices} 

In this Appendix we detail how we construct the resistance matrices between a pair of particles, $\Rpairlub$, and between a single particle and a bottom wall, $\Rwalllub$, for small inter-surface gaps. In both cases a combination of asymptotic formulas and tabulated numerical calculations are used to provide an accurate characterization of the resistance matrices across a wide range of dimensionless gap sizes; from very small to intermediate. 

\subsection{Semi-Analytical formulas for $\Rpairlub$}

There have been many works which calculate or tabulate the coefficients of $\Rpairlub$ for different particle separations $\epsilon_r$ \cite{SpheresStokes_JeffreyOnishi,LubricationLamb_3spheres,townsend2018generating}. Unfortunately we have found that no one of them provides sufficient accuracy at all distances we may wish to consider. Hence, we will use different formula for the coefficients $X\left(\epsilon_r\right)$ and $Y\left(\epsilon_r\right)$ appearing in $\Rpairlub$ (see eq. \eqref{Rpair}) depending on whether $\epsilon_r$ is small, large or some intermediate distance. We determine the cutoff distances for the different formula as the distances which minimize the error between formulas in neighboring regions, i.e., where the formulas `overlap'. 

At very small distances, Townsend gives asymptotic formulas for the coefficients of $\Rpairlub$ in \cite{townsend2018generating}. These asymptotic formulas break down as the particle separation is increased, and hence for large particle separations, we use tabulated values and linear interpolation for the coefficients of $\Rpairlub$ computed using Jeffrey and Onishi's series expansion \cite{SpheresStokes_JeffreyOnishi} truncated at 200 terms \footnote{We thank James Swan for providing us a Mathematica notebook which calculates this expansion.}. The mismatch between Townsend's asymptotic formulas and Jeffrey and Onishi's series formulas, however, is too large for all particle separations. In the interstitial region where neither Townsend nor Jeffrey and Onishi's suffice, we use tabulated values from Wilson's Fortran code \cite{LubricationLamb_3spheres} (based on Lamb's method of reflections) and linear interpolation to compute the coefficients of $\Rpairlub$. Minimizing the error between successive formulas gives the cutoff transitions: \newline
\begin{table}[h]
\centering
\begin{tabular}{cc}
Townsend: & $\epsilon_r < 6\times10^{-3}$ \\
Wilson: & $6\times10^{-3} < \epsilon_r < 10^{-1}$ \\
Jeffrey and Onishi: & $\epsilon_r > 10^{-1}$ \\
\end{tabular}
\end{table}

\subsection{Semi-Analytical formulas for $\Rwalllub$} \label{sec:rwall}  

For small wall--particle separations, we assemble asymptotic formulas for the coefficients from a few different sources. For larger wall--particle separations, we compute the coefficients using linear interpolation of tabulated values computed using our rigid multiblob method \cite{RigidMultiblobs} with 2562 blobs. For each coefficient we determine a cutoff transition distance for $\epsilon_h$ by minimizing the error between the asymptotic formulas and the multiblob values. 

\begin{figure}
\includegraphics[width=1\textwidth]{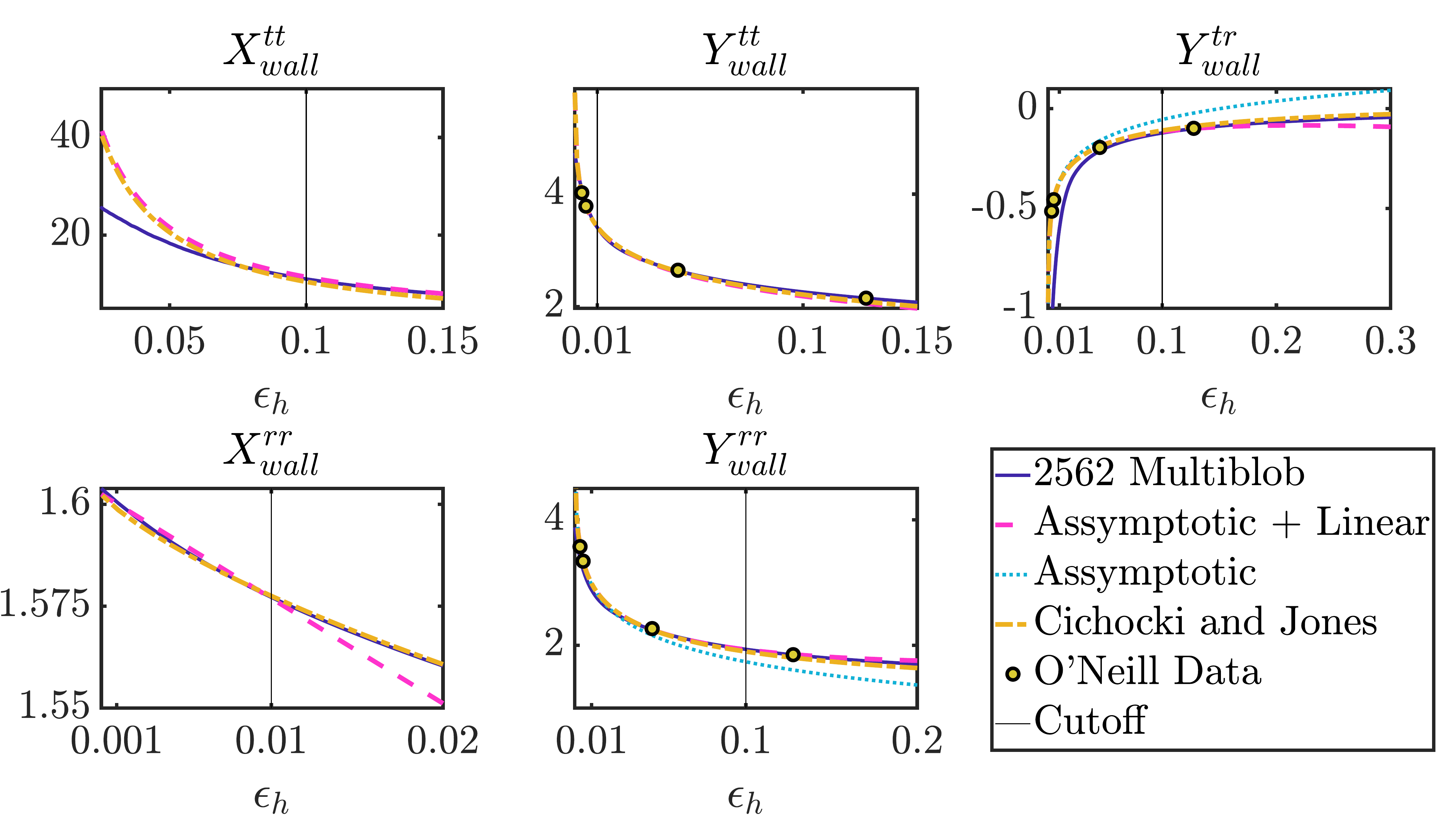}
\caption{Coefficients of $\Rwalllub$ appearing in equation \eqref{RwallXY} as a function the normalized gap $\epsilon_h$. Each panels shows that value of one of the coefficients computed using the rigid multiblob method \cite{RigidMultiblobs} with 2562 blobs (solid blue line), and the corresponding asymptotic result from table \ref{tab:Rwall} (dashed \modified{pink} line). Plots for $Y^{tr}_{\text{wall}}$ and $Y^{rr}_{\text{wall}}$ include the original asymptotic results computed in \cite{NearWallSphereMobility} (dotted \modified{green} line) as well as our modification to include a linear term \modified{(where appropriate)}. \modified{The 9/9 Pad\'e approximation given by Cichocki and Jones \cite{SphericalHarmonicImages_Wall} (dot--dashed yellow line) agrees well with out `Assymptotic+Linear' results within the cutoff regions.} Plots for $Y^{tt}_{\text{wall}}$, $Y^{tr}_{\text{wall}}$, and $Y^{rr}_{\text{wall}}$ also show data calculated by O'Neill (circles) and compiled in tables 1.1 and 2 respectively in \cite{NearWallSphereMobility}. The vertical black lines show the cuttoff transitions between different estimates.} 
\label{fig:wallmob}
\end{figure}

When $\epsilon_h$ is very small, Cooley and O'Neill give an asymptotic formula for $X^{tt}_{\text{wall}}\left(\epsilon_h \right)$ as equation (5.13) in \cite{SphereWall_Lubrication_ONeil}. This formula agrees with our multiblob computations for $\epsilon_h>0.1$ and hence we will take this as our cutoff value for this coefficient. Goldman, Cox, and Brenner give asymptotic formula for $Y^{tt}_{\text{wall}}\left(\epsilon_h \right),Y^{tr}_{\text{wall}}\left(\epsilon_h \right),Y^{rr}_{\text{wall}}\left(\epsilon_h \right)$ as equations (2.65a,b) and (3.13b) respectively in \cite{NearWallSphereMobility}. While these equations are certainly accurate enough at very small $\epsilon_h$, not enough terms are included in equations (2.65b) and (3.13b) to give good agreement with our multiblob results at larger $\epsilon_h$, or to give good agreement with the data provided by O'Neill in Table 1 of \cite{NearWallSphereMobility}. To remedy this, we add a linear term in $\epsilon_h$ to equations (2.65b) and (3.13b) from \cite{NearWallSphereMobility} and fit the coefficient to our multiblob results. Figure \ref{fig:wallmob} shows all of the coefficients of the wall mobility computed by combining the rigid multiblob method with asymptotic formulas. We see that for larger values of $\epsilon_h$, the new formulas we computed for $Y^{tr}_{\text{wall}}$ and $ Y^{rr}_{\text{wall}}$, which include a linear term in $\epsilon_h$, agree well with both our multiblob calculations as well as the data of O'Neill. Finally, an asymptotic formula for $X^{rr}_{\text{wall}}$ is given by Liu and Prosperetti in equation (4.1) of \cite{SphereWall_Lubrication_Spinner}. This formula largely agrees with out multiblob results for $\epsilon_h > 0.01$ so we use this as the cutoff. Table \ref{tab:Rwall} show the asymptotic formulas for each coefficient of $\Rwalllub$ (normalized by $1/(6 \pi \eta a)$) along with their respective cutoff values and sources. 

\modified{During the review process, an anonymous reviewer brought to our attention the work of Cichocki and Jones \cite{SphericalHarmonicImages_Wall}, which also computes the functions in table \ref{tab:Rwall}. They do this by taking the asymptotic formulas from Jeffrey and Onishi  \cite{SpheresStokes_JeffreyOnishi} for unequal spheres and taking the limit as one sphere becomes infinitely larger than the other one, and then complementing the asymptotics with a series expansion using images of vector spherical harmonics for larger gaps. Cichocki and Jones compute a 9/9 Pad\'e (rational) approximation for the difference between the (truncated) series expansion result and the asymptotics, see Eq. (63,43) in \cite{SphericalHarmonicImages_Wall}. We include some of these semi-analytical results in table \ref{tab:Rwall} and in Fig. \ref{fig:wallmob} for comparison with ours. We see that the Cichocki-Jones formula agrees with ours well over the range of gaps shown in the figure. However, for larger gaps we find a persistent difference with respect to our numerical results, which is significantly larger than the accuracy of the 2562-blob estimates. Furthermore, we find that some of the rational approximations exhibit poles for relative gaps of $O(1)$, and therefore are not sufficiently robust and accurate over the whole range of gaps we need in this work.}

\begin{table}[h!]
  \begin{center}
    \caption{Asymptotic formulas for the coefficients of $\Rwalllub$, along with their cutoff values and sources. \modified{The coefficient $0.95(88,43)$ in $Y^{tt}_{\text{wall}}$ indicates that the value we used was $0.9588$ while Cichocki and Jones \cite{SphericalHarmonicImages_Wall} give a value of $0.9543$; the two references agree for the coefficient $0.9713$ in $X^{tt}_{\text{wall}}$. For the coefficient $X^{rr}_{\text{wall}}$, the constant term matches in the two references, but the linear term is replaced by a term of $O\left(\epsilon_h\ln(\epsilon_h)\right)$ in Cichocki and Jones \cite{SphericalHarmonicImages_Wall}. For the remaining coefficients we add a linear term and fit the value of the constant together with the linear term, so direct comparison to Cichocki and Jones \cite{SphericalHarmonicImages_Wall} is not possible; see Fig. \ref{fig:wallmob} for a visual comparison.}}
    \label{tab:Rwall}
    \begin{tabular}{c|c|c|c} 
      \textbf{Coefficient} & \textbf{Formula} & \textbf{Cutoff} & \textbf{Source}\\
      \hline
      $X^{tt}_{\text{wall}}\left(\epsilon_h \right)$ & $\frac{1}{\epsilon_h} - \frac{1}{5}\log(\epsilon_h) + 0.9713$ & $\epsilon_h<0.1$ & (5.13) in \cite{SphereWall_Lubrication_ONeil}\\
      $Y^{tt}_{\text{wall}}\left(\epsilon_h \right)$ & $-\frac{8}{15} \log(\epsilon_h) + \modified{0.95(88,43)}$ & $\epsilon_h<0.01$ & (2.65a) in \cite{NearWallSphereMobility}\\
      $Y^{tr}_{\text{wall}}\left(\epsilon_h \right)$ & $\frac{4}{3} \left( \frac{1}{10}\log(\epsilon_h)+ 0.1895 - 0.4576 \epsilon_h \right)$ & $\epsilon_h<0.1$ & (2.65b) in  \cite{NearWallSphereMobility}  + linear\\
      $X^{rr}_{\text{wall}}\left(\epsilon_h \right)$ & $\frac{4}{3} \left( 1.2021 - 3(\frac{\pi^2}{6}-1) \epsilon_h \right)$ & $\epsilon_h<0.01$ & (4.1) in \cite{SphereWall_Lubrication_Spinner}\\
      $Y^{rr}_{\text{wall}}\left(\epsilon_h \right)$ & $\frac{4}{3} \left(-\frac{2}{5} \log(\epsilon_h) + 0.3817 + 1.4578 \epsilon_h \right)$ & $\epsilon_h<0.1$ & (3.13b) in \cite{NearWallSphereMobility} + linear\\
    \end{tabular}
  \end{center}
\end{table}

\section{\label{app:ValidationMultiblob}Accuracy of the lubrication approximation}

In this section we will asess the accuracy of the lubrication-corrected mobility $\epsM$, using the rigid multiblob method as a basis of comparison \cite{RigidMultiblobs,BrownianMultiblobSuspensions}. The multiblob method we use here does not include lubrication corrections but the accuracy can be improved by adding more blobs (nodes) per sphere.

\subsection{\label{app:Tetrahedron}Colloidal Tetrahedron}

\begin{figure}
\includegraphics[width=1\textwidth]{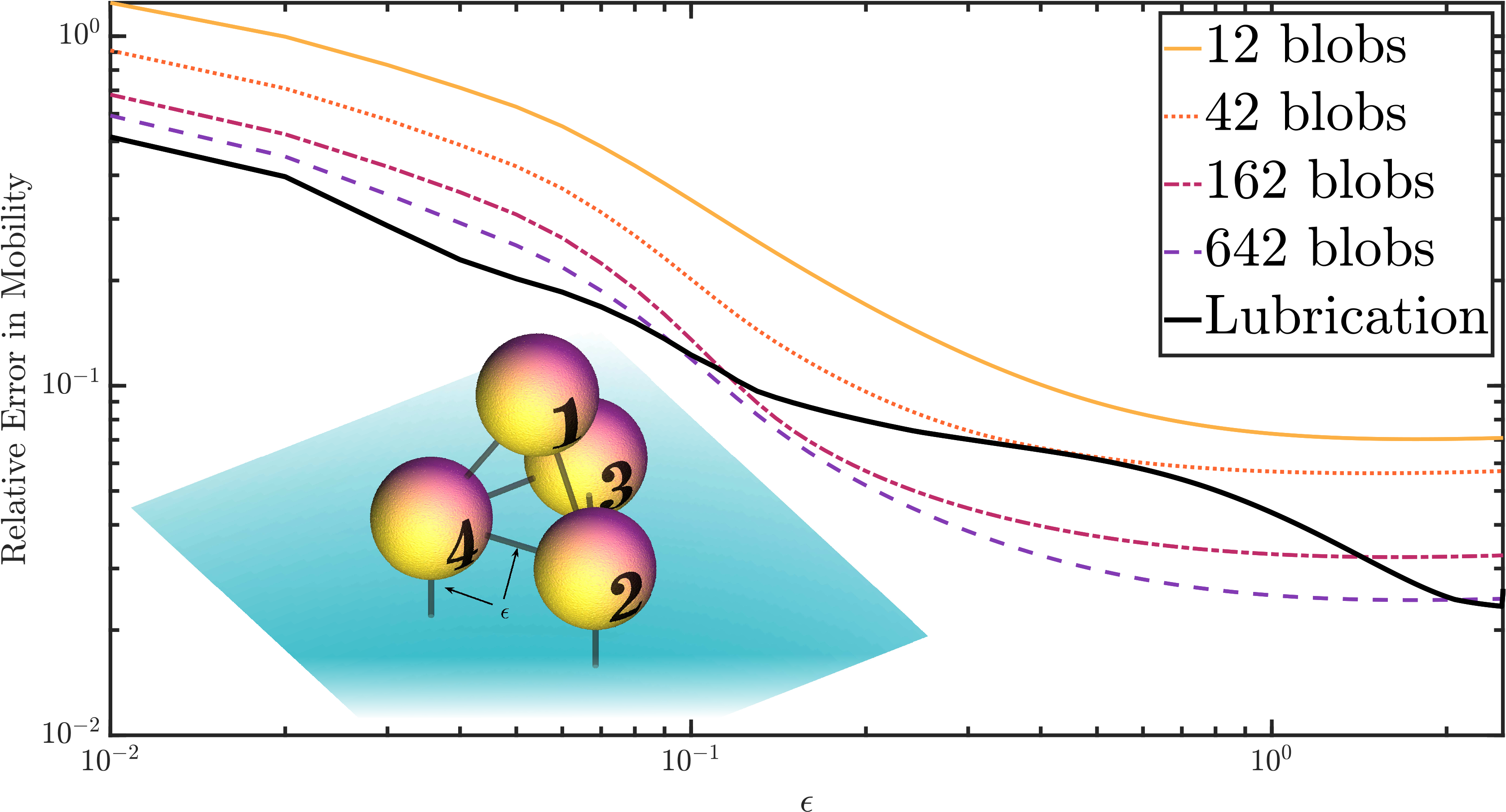}\caption{\label{fig:tetmob}Relative $L_2$ error in the mobility matrix of the colloidal tetrahedron shown in the inset as a function of the relative gap $\epsilon$, for the lubrication-corrected mobility $\epsM$ (solid black line), as well as the mobility matrix computed using the rigid multiblob method \cite{RigidMultiblobs} using 12, 42, 162, and 642 blobs to discretize each sphere. The error is measured relative to the mobility matix computed using 2562 blobs to discretize each sphere.}
\end{figure}

We first consider a colloidal tetrahedron above a wall, as depicted in the inset of Figure \ref{fig:tetmob}. Nearby particle surfaces are separated from each other by a distance $\epsilon$, which we vary $\epsilon$ as a control parameter. We compare the lubrication-corrected mobility $\epsM$ to that computed by the rigid multiblob method, for several different spatial resolutions. We use 12, 42, 162 and 642 blobs to discretize each sphere in the colloidal tetrahedron with the rigid multiblob method, and we take a calculation using 2562 blobs to be sufficiently accurate to provide a reference result \cite{RigidMultiblobs}.

Figure \ref{fig:tetmob} shows the relative error between the hydrodynamic mobility computed using the rigid multiblob method for different resolutions, as well as the lubrication-corrected mobility $\epsM$, as measured against our reference result. We see that for small $\epsilon$, the lubrication-corrected mobility is roughly as precise as the most accurate multiblob results and remains more accurate than both the 12 and 42 blob results for all distances considered. The error in $\epsM$ is larger than the more resolved multiblobs for intermediate separation distances $0.1 \lesssim \epsilon \lesssim 2$, but decays to approximately that of the 642-blob calculation for large values of $\epsilon$.

\subsection{\label{app:MB_Stoch}Dense Suspension of Microrollers}

Next we compare the particle displacements computed by the STS scheme summarized in Algorithm \ref{alg:trap} with those computed by the Trapezoidal Slip (TS) scheme developed for the rigid multiblob method in \cite{BrownianMultiblobSuspensions}. Specifically, we use both schemes to simulate the dense microroller suspension of $N = 2048$ particles studied in section \ref{sec:MagRoll}. We drive the suspension using a constant torque $\V \tau = 8 \pi \eta a^3 \omega \hat{\V{y}}$. The TS scheme, like the STS scheme, is a stochastic temporal integration method based on the deterministic trapezoid rule and we expect the two schemes to have similar temporal accuracy. Therefore we use a use a \emph{single step} of the STS and TS schemes with $\D{t} = 0.01$ to compute the one-step apparent velocities $V_x$ (i.e., particle displacements $V_x\D{t}$) along the direction of collective motion, and compare the results.

We use the distribution of one-step velocities, $P(V_x)$, computed by the STS scheme with lubrication corrections as a reference result, and compare with the TS scheme using $12$ and $42$ blobs per particle, without any lubrication corrections. To enable a direct comparison of the methods, we generate 100 statistically independent configurations at steady state using the STS scheme, and compute one-step apparent velocities starting from these configurations using the TS scheme with $12$ and $42$ blobs per particle. It is worthwhile noting that the lubrication-corrected BD method is not only considerably simpler but it is also more efficient; for our GPU-based implementation, one step of the TS scheme using $12$ blobs per sphere takes about 6 times longer, while using $42$ blobs per sphere takes almost 100 times longer, than one step of the STS scheme.

\begin{figure}
\includegraphics[width=1\textwidth]{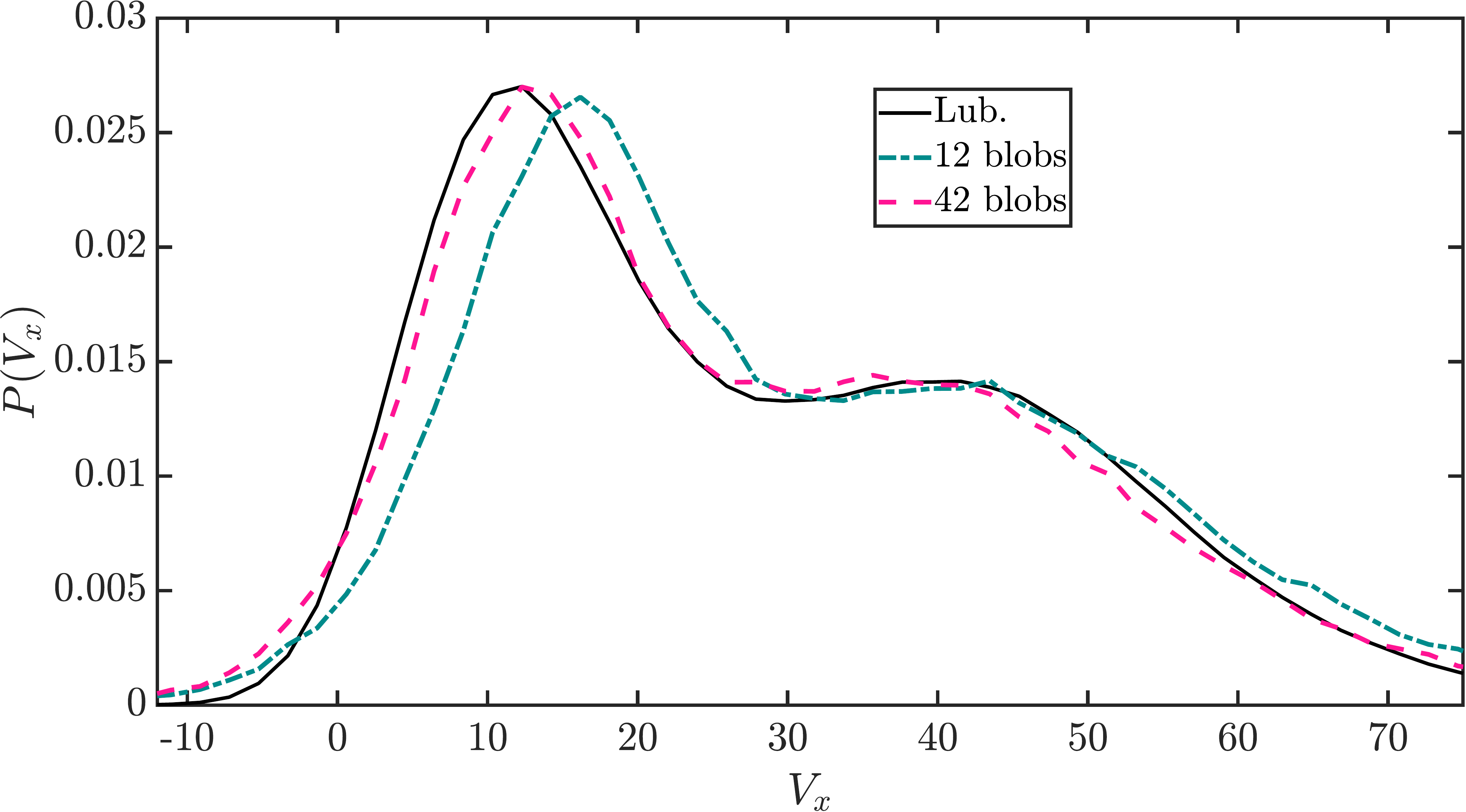}
\caption{\label{fig:MBcf}Histogram of one-step velocities for a dense uniform suspension of microrollers, computed using the lubrication-corrected STS scheme developed here (solid line), and the TS scheme of \cite{BrownianMultiblobSuspensions} using $12$ (dashed-dotted) and $42$ (dashed) blobs to discretize each sphere in the suspension.}
\end{figure}

Figure \ref{fig:MBcf} shows that the $P(V_x)$ distribution computed using the TS scheme approaches the distribution computed using the STS scheme as the spatial resolution of the TS scheme is increased from $12$ to $42$ blobs. The largest mismatch between the more accurate $42$ blob case and the lubrication-corrected BD method is the smallest velocities. We showed in section \ref{sec:MagRoll} that this is precisely the portion of the distribution due to particles nearest to the wall, and therefore most affected by lubrication. This example demonstrates that the minimally-resolved lubrication-corrected calculation is no less accurate overall than a 42-blob approximation that has not been corrected for lubricaton, as we already saw for the colloidal tetrahedron.

\section{Performance of Preconditioners} \label{sec:GMRES}

To interrogate the effectiveness of the preconditioner $\M{P}_{1}$, we consider a doubly--periodic suspension of $N_p$ spherical particles above a bottom wall. We take the particle radius $a=1 \mu$m and choose the particle's added mass $m_e$ to control the distribution of their height above the wall through the gravitational height $h_g - a = k_B T / (m_e g)  = 1/4 \mu$m, where $g$ is the acceleration of gravity. We change the in-plane packing fraction of the particles 
\begin{equation}
\phi = \frac{N_{p} \pi a^2}{L^2},
\end{equation}
where $L$ is the periodic length of the domain. Periodic boundary conditions are approximated using 8 periodic images as in \cite{MagneticRollers}.

Since the packing fraction is moderate compared to the theoretical in-plane packing limit ($\phi_{\text{max}}=\pi \sqrt{3}/6 \approx 0.91$), the particles form an approximate monolayer. Increasing $\phi$ can cause multi-layered particle configurations to become energetically favorable even for $\phi$ below the in-plane packing limit, due to the moderate gravitational height. In the remainder of this section, we will study how varying $\phi$ effects the convergence of the GMRES solver for \eqref{eq:LubSolve1} using both $\M{P}_{1}$ (see eq. \ref{eq:PC1}) and $\M{P}_{2}$ (see eq. \ref{eq:PCbd}) as preconditioners. We find that varying $h_g$ has only a mild effect on the convergence of the GMRES solver (not shown). 

\begin{figure}
\includegraphics[width=1\textwidth]{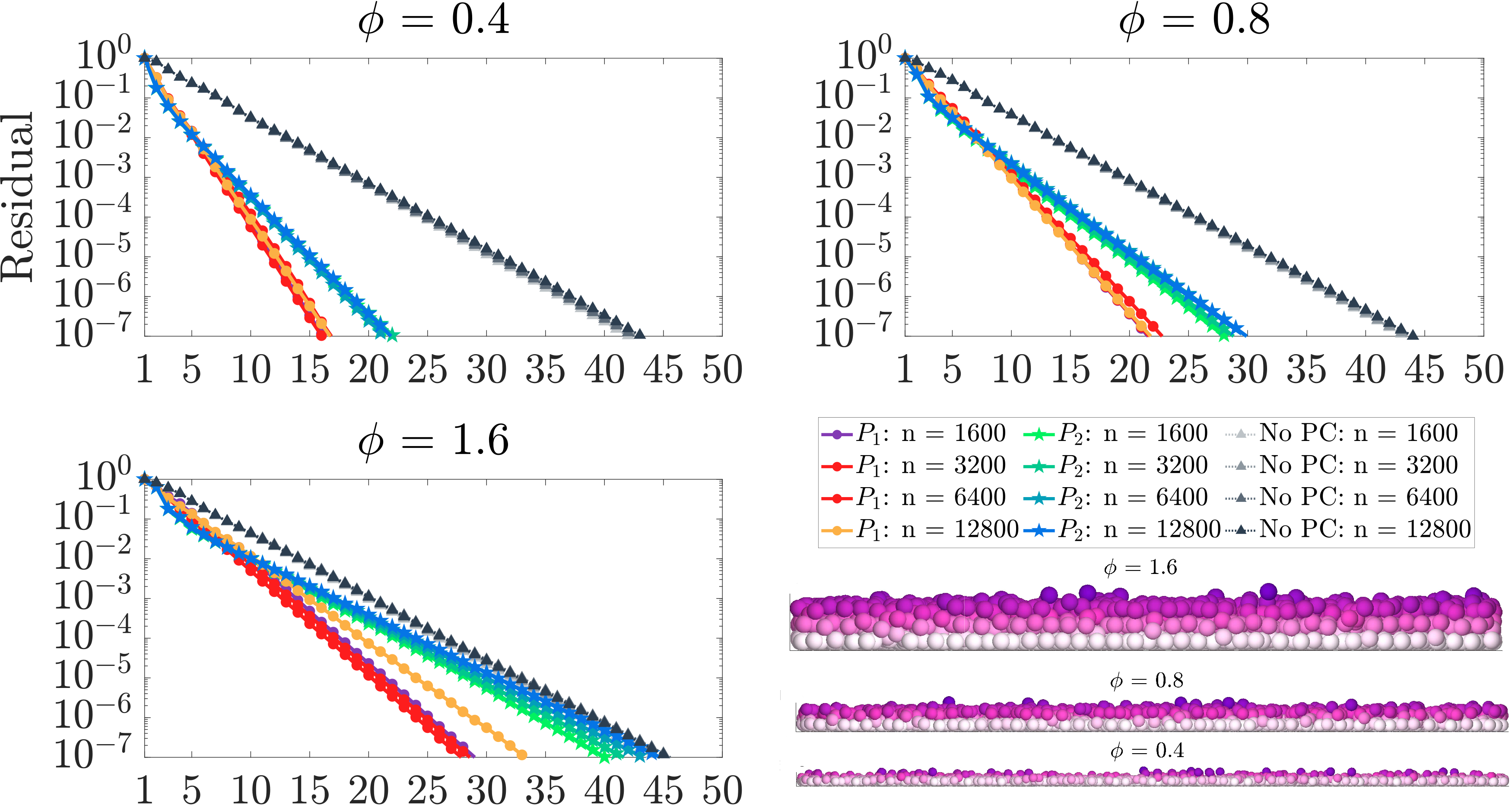}
\caption{Convergence rates of the GMRES solver for \eqref{eq:LubSolve1} using the proposed preconditioner $\M{P}_{1}$ , the block diagonal preconditioner $\M{P}_{2}$ \cite{SE_Multiblob_SD}, as well as an unpreconditioned GMRES method for reference (termed `No PC' in the legend). Each panel shows convergence rates for a fixed value of $\phi$, as the number of particles $N_p$ is varied. Below the legend is a frontal view of the particle configurations for $N_p=3200$ and for each value of $\phi$. Particles are colored based on their height above the wall with the highest particles colored the darkest while the lowest particles are colored the lightest. Higher values of $\phi$ cause multiple layers of particles to form.}
\label{fig:gmres}
\end{figure}

For $\phi = 0.4,0.8,1.6$, we increase the number of particles $N_p$ while keeping $\phi$ fixed. The reference configurations shown in Figure \ref{fig:gmres} illustrate how increasing $\phi$ increases the number of particle layers in the configuration from one for $\phi=0.4$ to about three at $\phi=1.6$. Figure \ref{fig:gmres} shows clearly that the preconditioner $\M{P}_{1}$ greatly improves the convergence of the GMRES solver over an unpreconditioned method for all of the values of $\phi$ considered. Further, the performance of the preconditioner is largely independent of $N_p$. 

The preconditioner $\M{P}_{2}$ performs similarly to $\M{P}_{1}$ for $\phi = 0.4,0.8$, but with a notably worse convergence rate for tighter tolerances ($<10^{-1}$) and more variation in the performance for different particle numbers. For $\phi=1.6$ the preconditioner $\M{P}_{2}$ performs only nominally better than no preconditioner at all, while $\M{P}_{1}$ gives some increased convergence; though not as much as the $\phi = 0.4,0.8$ cases. We suspect that $\M{P}_{1}$ outperforms $\M{P}_{2}$ in the multilayered case ($\phi=1.6$) because pairwise information is used to approximate $\Mob$ in $\M{P}_{1}$ but not in $\M{P}_{2}$. Clearly, however, multiple, tightly-packed layers of particles can hinder the effectiveness of both $\M{P}_{1}$ and $\M{P}_{2}$ as preconditioners. 

We note that the unpreconditioned method converges with roughly the same rate for each packing fraction $\phi$, gravitational height $h_g$, and all of the values of $N_p$ considered in each case. This is likely due to the hydrodynamic screening provided by the bottom wall which causes the hydrodynamic interactions between particles to decay like $1/r^3$ and aids in the conditioning of the mobility matrix \cite{MagneticRollers,BrownianMultiblobSuspensions}. Hence the presence of a bottom wall allows for an unpreconditioned GMRES method to be used while maintaining an overall complexity which scales linearly in the number of particles. Still, both preconditioners $\M{P}_{1}$ and $\M{P}_{2}$ are cheap, easy to compute and apply, and potentially speed up convergence by a factor of two to three; and therefore we employ $\M{P}_{1}$ in this work.      

\section{Weak accuracy of the STS scheme} \label{sec:proof}

In this appendix, we will prove that the value of $\V{U}^{n+1,\star}$ computed in step \ref{trap:7} of Algorithm \ref{alg:trap} is such that
\begin{equation} 
\V{U}^{n+1,\star} \eqd \epsM^{n+1,\star}  \V F^{n+1,\star} + 2 k_{B}T \left( \partial_{\V Q} \cdot \epsM \right)^{n} + \sqrt{\frac{2 k_{B}T}{\D{t}}}  \left(\epsM^{n}\right)^{1/2} \V W_{1,2} + \mathcal{R}\left(\D{t},\D{t}^{1/2}\right),
\end{equation}
where $\mathcal{R}\left(\alpha,\beta^{1/2}\right)$ denotes a Gaussian random error term with mean $\mathcal{O}\left(\alpha\right)$ and variance $\mathcal{O}\left(\beta\right)$. This combined with the fact that the predicted velocity computed in step \ref{trap:1} can be simplified using the shorthand notation from equations \eqref{eq:Mh1} and \eqref{eq:Mh2} as
\begin{align}
\V{U}^{n} \eqd \epsM^n \V F^{n} + \sqrt{\frac{2 k_{B}T}{\D{t}}} \left( \epsM^n \right)^{1/2} \V W_{1,2},  \label{eq:Upf}
\end{align}
shows that 
\begin{align}
 \Delta \V Q^{n+1} &= \V Q^{n+1}-\V Q^{n} = \frac{\D{t}}{2} \left( \V U^{n} + \V{U}^{n+1,\star} \right) \\
 &\eqd \frac{\D{t}}{2} \left( \epsM^n \V F^{n} + \epsM^{n+1,\star}  \V F^{n+1,\star} \right) \\
 &+ k_{B}T \D{t} \left( \partial_{\V Q} \cdot \epsM \right)^{n} + \sqrt{2 k_{B}T \D{t}}  \left(\epsM^{n}\right)^{1/2} \V W_{1,2} + \D{t} \ \mathcal{R}\left(\D{t},\D{t}^{1/2}\right). \label{eq:DesiredDrift}
\end{align}
This proves that the STS scheme obtains the correct stochastic drift, is second-order accurate in the deterministic setting, and is weakly first-order accurate in the stochastic setting.

For simplicity, we take $\V F \equiv \V 0$ at all time levels as the main difficulty here is showing that the stochastic increments are correct. Using the RFD approximation \eqref{eq:RFD}, we write the value of $\V{U}^{n+1,\star}$ computed in step \ref{trap:7} (with $\V F \equiv \V 0$) as 
\begin{align} \label{eq:UcorRFD}
\D{t}\,\V{U}^{n+1,\star} &\eqd 2  k_{B}T \D{t} \left[\M{I} + \Mob^{n+1,\star} \Lub^{n+1,\star} \right]^{-1} \left( \partial_{\V Q} \cdot \epsM \right)^{n} \\
& \hspace{-1cm} + \sqrt{2 k_{B}T\D{t}}  \left[\M{I} + \Mob^{n+1,\star} \Lub^{n+1,\star} \right]^{-1}  \left(\Mob^n + \Mob^n \Lub^n \Mob^n \right)^{1/2} \V W_{1,2} + \mathcal{R}\left(\delta^{2},\D{t}\right). \nonumber
\end{align}
Now if we Taylor expand $\left[\M{I} + \Mob^{n+1,\star} \Lub^{n+1,\star} \right]^{-1}$ about the configuration $\V{Q}^n$, we may write
\begin{align}
& \left[\M{I} + \Mob^{n+1,\star} \Lub^{n+1,\star} \right]^{-1} \left( \partial_{\V Q} \cdot \epsM \right)^{n} =\\
& \left[\M{I} + \Mob^{n} \Lub^{n} \right]^{-1} \left( \partial_{\V Q} \cdot \epsM \right)^{n} + \mathcal{R}\left(\D{t},\D{t}^{1/2}\right), \label{eq:UcorDQ}
\end{align}
where in the last equality we have used the fact that 
\begin{align} 
\Delta \V{Q}^{\star} &= \V{Q}^{n+1,\star} - \V{Q}^{n} = \D{t}\,\V{U}^{n} \nonumber \\
&\eqd \sqrt{2 k_{B}T \D{t}}  \left[\M{I} + \Mob^{n} \Lub^{n} \right]^{-1}  \left(\Mob^n + \Mob^n \Lub^n \Mob^n \right)^{1/2} \V W_{1,2} = \mathcal{R}\left(0,\D{t}^{1/2}\right). \label{eq:DeltaQP}
\end{align}
 
By Taylor expanding the second term in equation \eqref{eq:UcorRFD} around $\V{Q}^n$ and using the shorthand \eqref{eq:Mh2} and equation \eqref{eq:DeltaQP}, we may write
\begin{align}
&\sqrt{\frac{2 k_{B}T}{\D{t}}}  \left[\M{I} + \Mob^{n+1,\star} \Lub^{n+1,\star} \right]^{-1}  \left(\Mob^n + \Mob^n \Lub^n \Mob^n \right)^{1/2} \V W_{1,2} = \\
& \sqrt{\frac{2 k_{B}T}{\D{t}}} \left(\epsM^{n}\right)^{1/2} \V W_{1,2} + 2 k_{B}T  \left( \partial_{\V{Q}}  \left[\M{I} + \Mob \Lub \right]^{-1} \right)^{n} \colon \\ \nonumber
&\left[ \left(\Mob^n + \Mob^n \Lub^n \Mob^n \right) \left(\M{I} + \Mob^{n} \Lub^{n} \right)^{-T} \V W_{1,2} \V W_{1,2}^{T}  \right] + \mathcal{R}\left(\D{t}, \D{t}^{1/2} \right) = \\
&\sqrt{\frac{2 k_{B}T}{\D{t}}} \left(\epsM^{n}\right)^{1/2} \V W_{1,2} + 2 k_{B}T  \left( \partial_{\V{Q}}  \left[\M{I} + \Mob \Lub \right]^{-1} \right)^{n} \colon \Mob^{n}  + \mathcal{R}\left(\D{t}, \D{t}^{1/2} \right). \label{eq:UcorFin}
\end{align}
Combining equations \eqref{eq:UcorFin} and \eqref{eq:UcorDQ} with equation \eqref{eq:UcorRFD} and using equation \eqref{eq:splitRFD} from the main text to simplify 
\[
\left[\M I+\Mob\Lub\right]^{-1} \left( \partial_{\V{Q}} \cdot \Mob \right) + \left( \partial_{\V{Q}} \left[\M I+\Mob\Lub\right]^{-1} \right) \colon \Mob = \partial_{\V{Q}} \cdot \epsM,
\]
gives the desired result \eqref{eq:DesiredDrift}.

\section{Experimental details} \label{sec:exp_appendix}

\label{sec:exp_DLS}
For the SEM size measurement the particles were imaged using a Gemini Field Emission Scanning Electron Microscope (Zeiss).
In the DLS measurement the particles were dispersed in a nonionic density gradient medium \cite{Rickwood1982} mixed with water to prevent significant sedimentation during the measurement. Iohexol (Sigma-Aldrich) was mixed with ultrapure water (Milli-Q, Millipore) at a 74 w/v\% concentration (density: 1.39 g/mL) and the viscosity of the mixture was measured to be 17.2 cP (22 \textdegree C) using an Ubbelohde viscometer (CANNON Instrument Company). The DLS measurement was done using a Zetasizer Nano ZS (Malvern Instruments Ltd.).

\label{sec:exp_cell}
The glass sample cell was constructed in the following way: two glass spacers (No. 1 coverslips, $\sim$150 $\mu$m thick) were glued to a microscope slide with a $\sim$3 mm separation using UV glue (Norland Adhesives, No. 68). On top of this a basebath-treated coverslip was glued to created a channel. This channel was filled with the dispersion and both ends were glued shut. In the final step the UV glue was cured while the dispersion was shielded from the UV light by a piece of aluminum foil, to prevent the bleaching of the dye inside the particles. After curing, the sample was placed with the coverslip down.

\label{sec:exp_diff}
For the measurement of the diffusion constant $\bar{D}_{||}$ of the particles parallel to the glass wall, fluorescent particles were imaged with an inverted microscope (IX83, Olympus) and a 20$\times$/0.7 NA air objective in fluorescent mode with 488 nm LED excitation.

\begin{figure}
\includegraphics[width=1\textwidth]{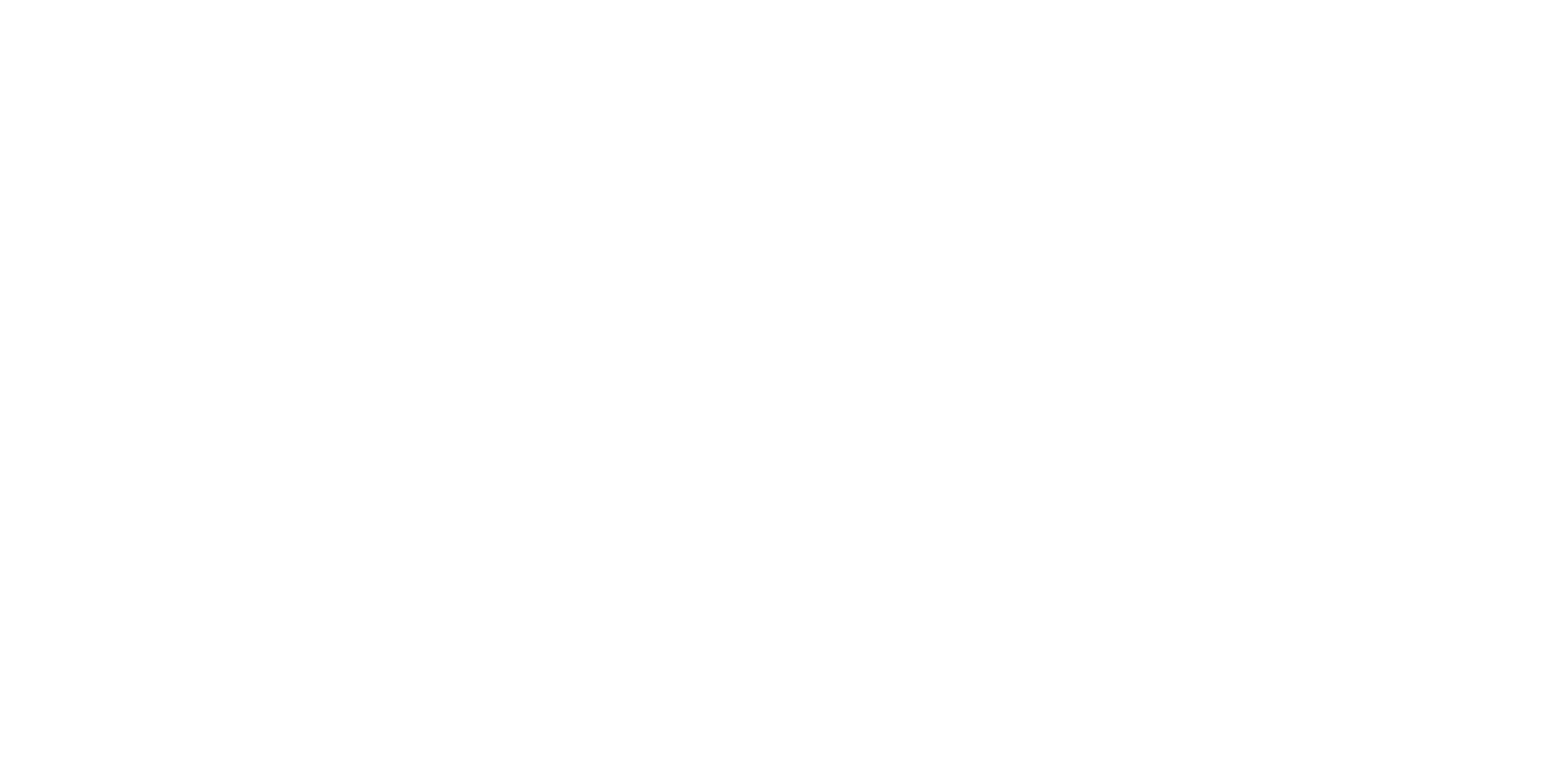}\caption{\modified{Photograph and schematic (side view) of the setup used for the roller experiments. The setup consists of a home-built tri-axial Helmholtz coil set \cite{Abbott2015} mounted on the stage of an inverted microscope. The sample is placed in the center of the coil set and the microscope objective is raised into the coil set using an extension tube.}}
\label{fig:expsetup}
\end{figure}

\label{sec:exp_setup}
For the roller experiments a home-built tri-axial nested Helmholtz coil set \cite{Abbott2015} was put on top of an inverted microscope (IX83, Olympus) to allow simultaneous imaging and magnetic field exposure. The square coil bobbins were made by 3D printing \modified{(see Figure~\ref{fig:expsetup}).} The sample was placed in the center of the coil set and an extension tube (Thorlabs) was used to raise the objective (20$\times$/0.7 NA air) into the center of the coil set. A $\pi/2$ out-of-phase sinusoidal magnetic field (40 G) was generated by two coils using a computer code, a data acquisition system (DAQ, Measurement Computing) and two AC amplifiers (EMB Professional). One of the two coils was parallel to gravity and the optical axis of the microscope, while the other was perpendicular to the first, resulting in a rotating magnetic field perpendicular to the lateral plane of imaging and bottom glass wall. The fluorescently labelled particles in the middle of the channel were imaged in fluorescent mode using 488 nm LED illumination at a frame rate of 9.0 s$^{-1}$. At the same time the particles were kept in focus using a drift compensation module (IX3-ZDC2, Olympus) in continuous mode. To prevent the particles from ending up at one side of the sample container, the direction of the rotating field was inverted every 30 seconds. 


\end{document}